\documentclass[sigconf]{acmart}
\usepackage[utf8]{inputenc}
\usepackage{booktabs}
\usepackage{tabularx}
\usepackage[table]{xcolor}
\usepackage[dvipsnames]{xcolor}
\usepackage{multirow}

\definecolor{hblue}{HTML}{D0E1F9}   
\definecolor{corange}{HTML}{FAD7AC} 
\definecolor{agreen}{HTML}{D1E8D1}  

\usepackage{tikz}
\newcommand*\blackcircled[1]{%
  \tikz[baseline=(char.base)]{
    \node[shape=circle,draw,fill=black,inner sep=0.2pt] (char) {\textcolor{white}{#1}};
  }
}

\AtBeginDocument{%
  }

\setcopyright{none} 
\copyrightyear{2026}
\acmYear{2026}
\acmDOI{XXXXXXX.XXXXXXX}

\acmConference[MICRO 2026]{The 58th IEEE/ACM International Symposium on Microarchitecture}{October 31--November 04, 2026}{Athens, Greece}

\acmISBN{978-X-XXXX-XXXX-X/XX/XX}

\begin{document}


\title{D-NOVA: In-Storage Retrieval Accelerator via Dual-Bound
\\ 3D NAND-Optimized Similarity Search with Vector Adaptation}

\author{Chang Eun Song}
\authornote{These authors contributed equally to this work.}
\affiliation{
  \institution{University of California, San Diego}
  \city{La Jolla}
  \country{USA}
}
\email{cesong@ucsd.edu}

\author{Sumukh Pinge}
\authornotemark[1]
\affiliation{
  \institution{University of California, San Diego}
  \city{La Jolla}
  \country{USA}
}
\email{spinge@ucsd.edu}

\author{Tianqi Zhang}
\affiliation{
  \institution{University of California, San Diego}
  \city{La Jolla}
  \country{USA}
}
\email{tiz014@ucsd.edu}

\author{Sung Eun Kim}
\affiliation{
  \institution{University of California, San Diego}
  \city{La Jolla}
  \country{USA}
}
\email{suk062@ucsd.edu}

\author{Tajana S Rosing}
\affiliation{
  \institution{University of California, San Diego}
  \city{La Jolla}
  \country{USA}
}
\email{tajana@ucsd.edu}

\author{Mingu Kang}
\affiliation{
  \institution{University of California, San Diego}
  \city{La Jolla}
  \country{USA}
}
\email{mingu@ucsd.edu}


\begin{abstract}
Retrieval-Augmented Generation (RAG) enhances the factual grounding of large language model (LLM) inference by retrieving 
relevant information from external knowledge bases. However, its dense vector retrieval introduces significant latency and energy overhead, becoming the primary performance bottleneck. Although recent in-storage accelerators aim to reduce data movement, they still rely on host or embedded processors outside the memory, where nearly 70\% of the total retrieval time is spent. As a result, they cannot fully overcome the bandwidth limitations, leading to yet another memory bottleneck.
To tackle these limitations, we present D-NOVA, a hardware–software co-designed in-storage retrieval accelerator. D-NOVA executes an inverted file (IVF)-based hierarchical retrieval pipeline by deeply embedding the search functionality directly into the NAND memory array. This is achieved by incorporating a new distance metric, Dual-bound Tight Similarity Sensing (DTS), which is specifically tailored for searching within the NAND string.
In addition, we introduce a lightweight contrastive adapter that maps embedding vectors into a DTS–friendly domain, recovering near-software recall while improving performance and energy efficiency. D-NOVA is up to 41.7× faster and 71× more energy-efficient than a CPU baseline, and achieves 12.13× higher throughput while being up to 1.26× more energy-efficient than state-of-the-art in-storage RAG accelerators, demonstrating the potential of fully in-storage vector search for scalable RAG acceleration.
\end{abstract}
\vspace{-0.35cm}



\maketitle

\section{Introduction}

\begin{figure}[t]
    \centering
    \includegraphics[width=0.85\linewidth]{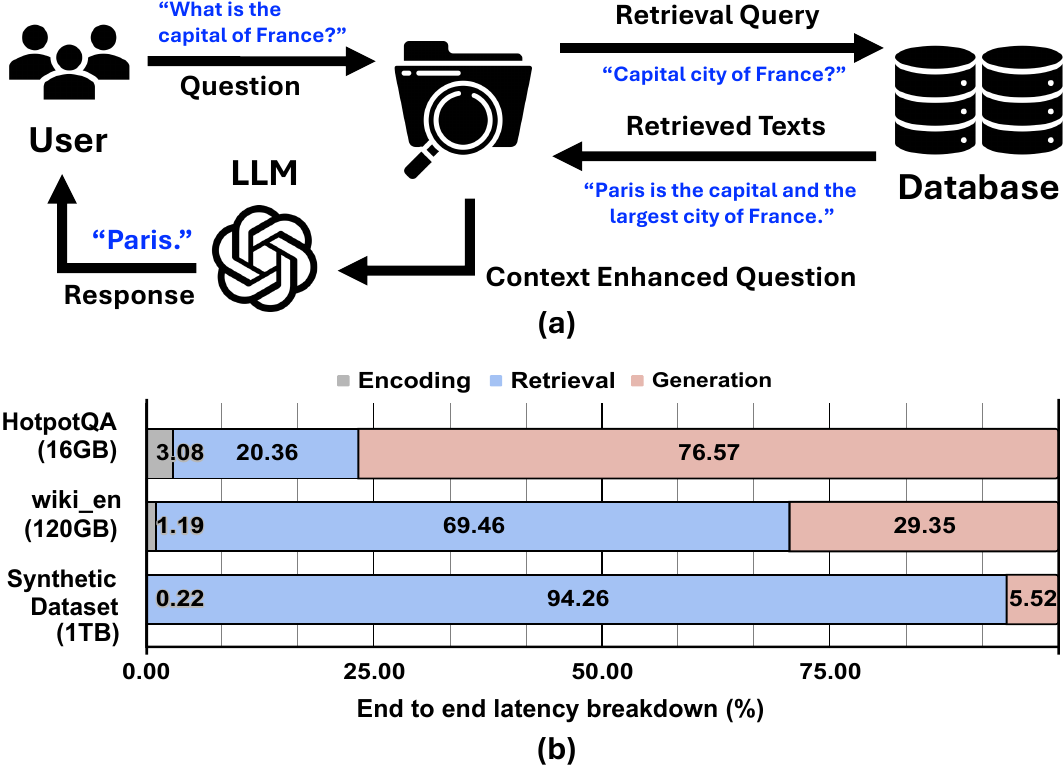}
    \caption{(a) RAG system overview, and (b) end-to-end latency breakdown of IVF-Flat with~\cite{cohere_wikipedia_2023_11_embed_multilingual_v3},~\cite{hotpotdataset}, and synthetic datasets.}
    \label{fig:intro}
\end{figure}
 
Large language models (LLMs) have revolutionized language understanding and, through recent extensions, enabled reasoning across multiple modalities. Their inference increasingly relies on GPUs and specialized hardware for acceleration \cite{vllm, 10.1145/3695053.3730999, 10.1145/3579371.3589057, yazdanbakhsh2022sparse, 10.1145/3665314.3670798, yang2024fslhdnn, song202452}. However, their knowledge remains largely static, constrained by pre-training data and costly to update. To mitigate this limitation, Retrieval-Augmented Generation (RAG) has emerged as an effective solution. As illustrated in Fig.~\ref{fig:intro}(a), 
RAG augments LLMs with external knowledge bases, grounding outputs in up-to-date information without retraining, but this benefit comes with a major retrieval bottleneck.
Even with algorithmic optimizations and hardware acceleration on CPUs and GPUs~\cite{cagra, hu2025hedrarag, jiang2025rago, quinn2025accelerating, jin2024ragcache}, each query must still traverse billions of high-dimensional embeddings distributed across the memory hierarchy from storage to DRAM and on to the processor~\cite{reis, 10.1145/3695053.3731032, hermes}. 
This massive data movement dominates latency and energy, offsetting much of the benefit of computational acceleration. 
As the vector database (DB) scales, empirical studies~\cite{hermes, reis, jeong2025call, 10.1145/3695053.3731032, wang2025turbocharging} show that retrieval alone can spend more than 70\% of the total end-to-end latency (Fig.~\ref{fig:intro}(b)), highlighting data movement as the main bottleneck in current RAG systems.

To address this data bandwidth limitation, approximate nearest neighbor (ANN) algorithms such as Hierarchical Navigable Small World (HNSW) graphs and Inverted File (IVF) indices are widely adopted for improving retrieval efficiency~\cite{reis,cagra,hnswSMARTSSD,11202651}. However, they impose very different demands on hardware design. HNSW achieves high recall accuracy, which comes at the cost of irregular pointer-chasing behavior and unpredictable access patterns~\cite{xu2023proxima, kim2022accelerating}, making it difficult to implement efficiently on highly parallel hardware platforms~\cite{cagra}. In contrast, IVF provides a more structured access pattern, making it better suited for parallel and hardware-efficient execution. 
Nevertheless, both HNSW and IVF remain fundamentally constrained by data movement overhead, as large volumes of vector data must still be transferred across the memory/storage hierarchy. This shared bottleneck has motivated the development of in- and near-storage architectures.

Near-storage retrieval accelerators~\cite{10.1145/3695053.3731032, hnswSMARTSSD} bring processing closer to SSDs, but still require substantial architectural modifications and depend on host CPUs or GPUs for synchronization and final re-ranking. For instance, RAGX~\cite{10.1145/3695053.3731032} and SmartSSD~\cite{hnswSMARTSSD,9141369} embed reconfigurable or FPGA-based accelerators near the storage device. However, their limited on-device memory capacity and strict power constraints hinder scalability beyond relatively lightweight data pipelines. REIS~\cite{reis} leverages in-storage processing (ISP) by applying binary quantization for coarse ANN search. However, due to its limited accuracy, it requires an expensive INT8 re-ranking that retrieves high-precision data from flash to the SSD controller and DRAM. This re-ranking step accounts for nearly 70\% of the total retrieval time, dominated by multi-level cell (MLC) read latency and data movement to the embedded processor, resulting in frequent data transfers between storage, controllers, and on-chip memory, which create a new memory wall that fundamentally limits bandwidth, energy efficiency, and parallelism in search.
A natural next step is to embed search operations deeper in the NAND array.
Unlike ReRAM or PCM crossbar arrays~\cite{song2025hybrid, fan2024specpcm}, which rely on analog current accumulation for dot-product computation, NAND flash inherently operates on discrete states, e.g., serial string connect vs. disconnect, making such analog crossbar-style accumulation impractical.
In particular, the compact cell with only one transistor makes it difficult to add any functionality, which greatly restricts the capability of the NAND-based ISP.

To address these shortcomings, we present D-NOVA (\textbf{\underline{D}}ual-Bound \textbf{\underline{N}}AND-based \textbf{\underline{O}}ptimized search with \textbf{\underline{V}}ector \textbf{\underline{A}}daptation). 
D-NOVA is a highly parallel in-storage retrieval architecture that embeds search functionality deeply into the memory array with negligible peripheral hardware changes.
It also incorporates a new distance metric, \textit{dual-bound tight similarity sensing} (DTS), tailored for in-array execution within NAND strings.
DTS reformulates vector retrieval as query-dependent threshold search over ordered $V_{\mathrm{TH}}$ levels, using complement sensing for accurate search and compact scoring.
Unlike prior in-memory computing approaches~\cite{song2025hybrid, fan2024specpcm} that often rely on analog current accumulation, D-NOVA is strictly digital and operates solely through binary connect/disconnect decisions along NAND strings, without any reliance on analog accumulation or current summation.
It also executes the three-stage IVF pipeline, including centroid search, coarse Top-$K_2$ retrieval, and fine Top-$K_1$ re-ranking, entirely within 3D NAND arrays, while the SSD controller handles only lightweight scoring operations. D-NOVA's key contributions are:

\noindent $\bullet$  \textbf{In-storage searching and retrieval:}
    D-NOVA performs true in-storage search within the NAND’s multi-level cells, where similarity is evaluated directly inside the NAND array by comparing the stored cell values with query-driven wordline voltages. It reuses existing peripheral circuitry to perform local group-wise comparisons with negligible hardware overhead. The operation is fully digital, relying on on/off evaluations that avoid the analog non-idealities observed in analog-domain processing.
    
\noindent $\bullet$  \textbf{Accurate DTS search:} 
    We propose a DTS framework that serves as a tailored distance metric for performing search operations in the NAND string. DTS narrows the candidate set by applying tight dual bounds (upper and lower) during search. 
    This enables fine-grained similarity evaluation using simple binary sensing instead of the iterative read process required for multi-level cell reads.

\noindent $\bullet$  \textbf{Database domain conversion through custom adapter:} 
    We introduce an adapter that projects embeddings into a DTS-aligned space, bridging the gap between widely used cosine/L2 metrics and DTS. The adapter can be trained once \textit{offline} using a contrastive objective on document vectors, restoring near-full IVF recall with near-zero runtime overhead for in-storage search.
    
\noindent $\bullet$  \textbf{Controllable efficiency and accuracy configuration:} 
    We provide a controllable knob to trade off efficiency and accuracy. The number of wordlines sensed in parallel per block, denoted as $m$, is dynamically adjusted across IVF stages, given their accuracy requirements and data volume for the stage-wise optimization.


Through this cross-layer co-design, D-NOVA demonstrates that large-scale semantic retrieval can be executed almost entirely in-storage, effectively eliminating data movement while maintaining high accuracy. 
When evaluated on the NQ and HotpotQA datasets, D-NOVA achieves $33.3-41.7\times$ and $4.3-12.1\times$ higher throughput and $58.8-71\times$ and $1.1-1.26\times$ lower energy consumption than CPU baseline and SOTA~\cite{reis}, respectively, establishing a new direction for scalable and energy-efficient in-storage retrieval acceleration. Its robustness and feasibility are also validated through detailed noise simulations (Section~\ref{alphasensitivity}), while implementation overheads, including area and power, are quantified in Section~\ref{sec:overhead}.

\begin{figure}[t]
    \centering
    \includegraphics[width=1\linewidth]{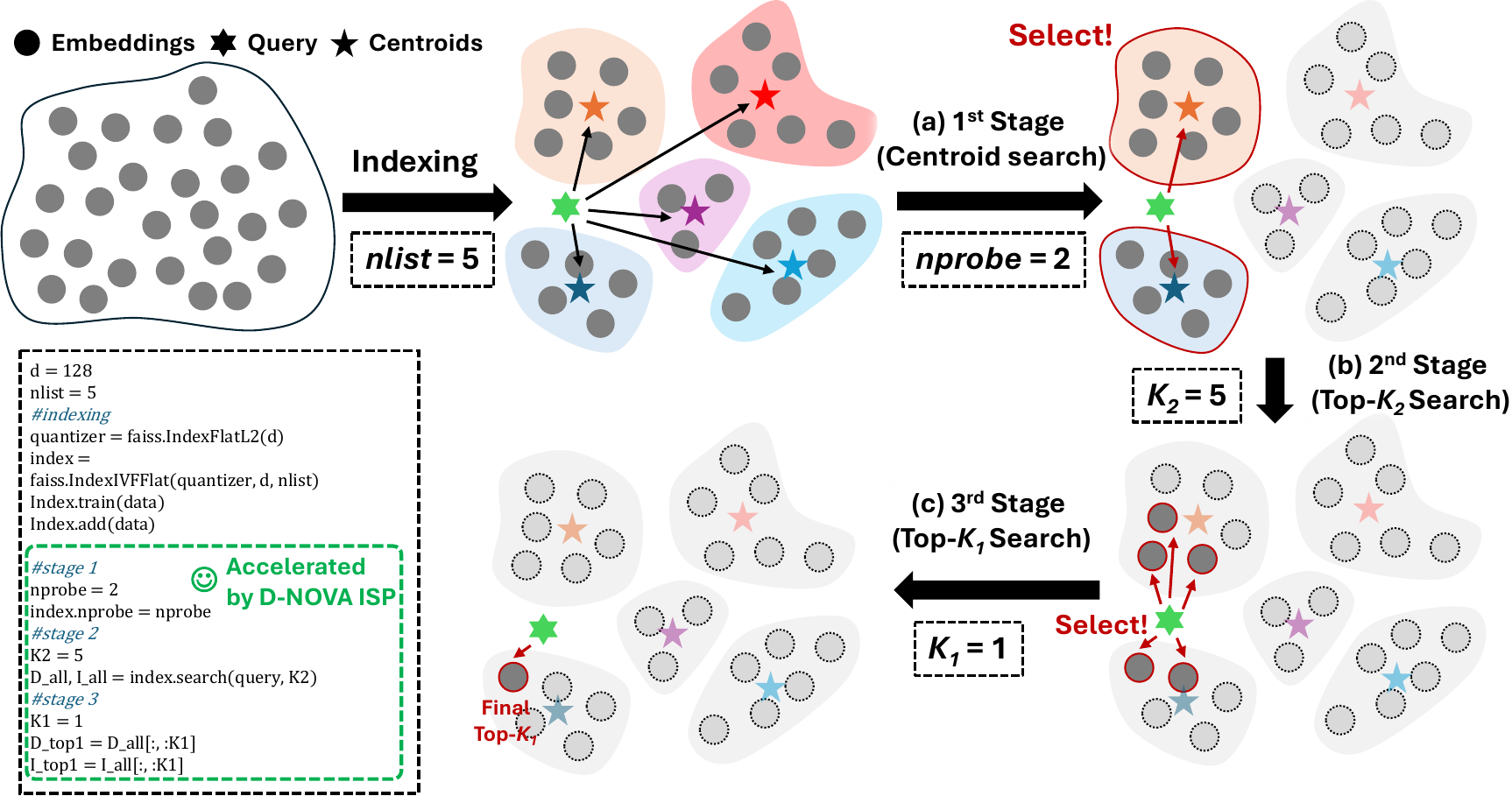}
    \caption{IVF three stages. (a) Centroid search, (b) coarse-grained Top-$K_2$ search, and (c) Top-$K_1$ search (Re-ranking)}
    \label{fig:ivf}
\end{figure}

\section{Background and Related Works}

\subsection{RAG Retrieval Pipeline and IVF Hierarchy}


RAG systems consist of two main stages: (1) \textit{retrieval}, which locates semantically relevant documents from a large embedding DB given the query, and (2) \textit{generation}, where an LLM uses the retrieved context to produce the final response. The retrieval stage dominates the system cost, as it involves searching billions of high-dimensional vectors per query~\cite{wang2020minilm, song2020mpnet}. This search is typically implemented using ANN algorithms, such as the IVF hierarchy~\cite{11202651, zobel2006inverted}, due to IVF's scalability and predictable access patterns.

The IVF hierarchy decomposes search into three coarse-to-fine stages (Fig.~\ref{fig:ivf}). 
First, all DB embedding vectors are partitioned into $N_{nlist}$ (e.g., $10^3$--$10^4$) clusters via $k$-means~\cite{ahmed2020k}, and the query $q$ is compared against all centroids to select the top $N_{nprobe}$ (e.g., $8$--$256$) clusters (Fig.~\ref{fig:ivf}(a)). 
Second, only vectors in the probed clusters are evaluated to produce a coarse Top-$K_2$ candidate list (typically $K_2 \approx 5$--$20\times K_1$) (Fig.~\ref{fig:ivf}(b)). 
Finally, a fine-grained re-ranking stage ranks the coarse Top-$K_2$ candidates and selects the Top-$K_1$ (e.g., $1$--$100$) results (Fig.~\ref{fig:ivf}(c)).
After the Top-$K_1$ indices are determined, the corresponding raw documents are fetched from storage to the LLM for generation. 
This structured, batched pipeline (centroid search $\rightarrow$ coarse-grained search $\rightarrow$ compact re-ranking) yields predictable and parallel access patterns, making IVF hardware-friendly.

Conventional software frameworks such as FAISS~\cite{11202651} typically store the IVF index in host memory~\cite{zobel2006inverted}, but execute different retrieval stages—centroid assignment, candidate scoring, and re-ranking—on heterogeneous compute engines such as CPUs and GPUs~\cite{fusionanns}, where CPUs handle control-intensive centroid assignment while GPUs accelerate large-scale similarity computations. 
As a result, embeddings and partial scores are frequently transferred between host DRAM and accelerator memory, and for billion-scale databases, these PCIe or NVLink transfers dominate both latency and energy~\cite{johnson2019billion,li2025scaling}.
Unlike graph-based methods such as HNSW, whose irregular pointer-chasing and unpredictable access patterns hinder efficient hardware mapping~\cite{xu2023proxima}, IVF offers structured and highly parallel access patterns, making it a strong candidate for ISP or near-storage processing (NSP) acceleration, where similarity computations can be executed directly inside or near the storage arrays~\cite{reis, hnswSMARTSSD}. However, if any pipeline stage is not covered by the ISP/NSP functionality, intermediate data must still leave the storage, reintroducing data movement as the dominant bottleneck.



\subsection{Limitations of Existing RAG Accelerators}

\begin{figure}[t]
    \centering
    \includegraphics[width=0.95\linewidth]{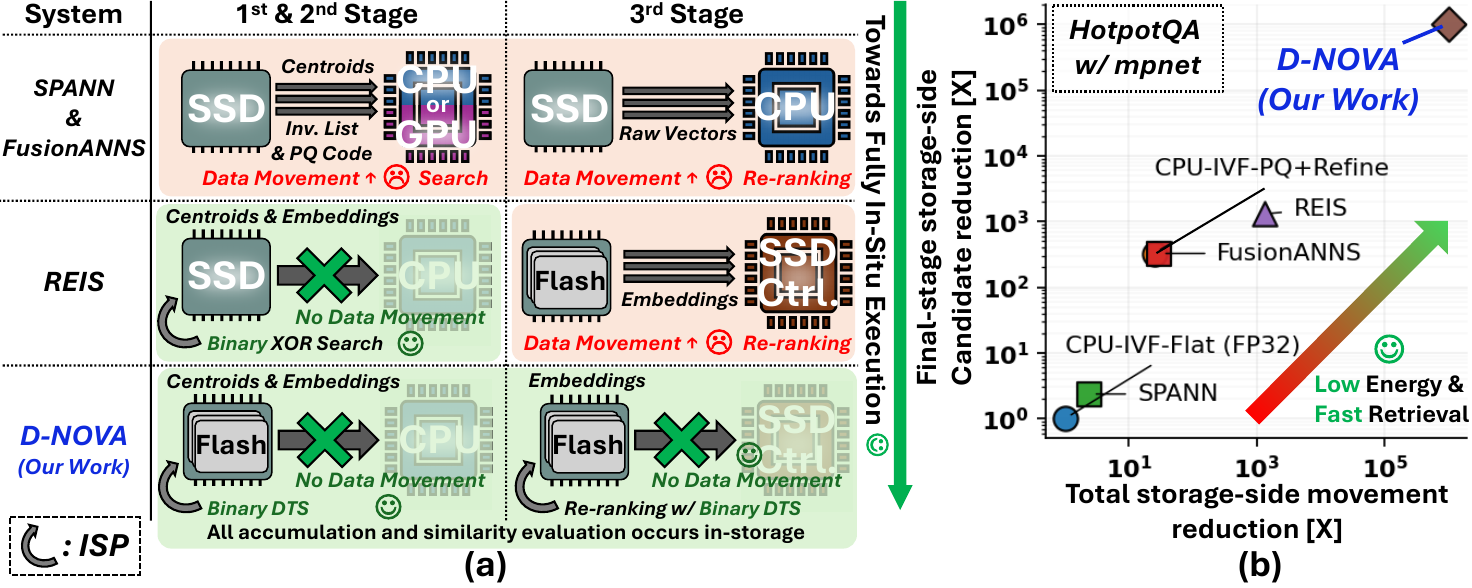}
    \caption{(a) Movement-centric comparison of representative retrieval designs, and (b) data movement reduction plot.}
    \label{fig:motiv}
\end{figure}

Several prior works explore executing the retrieval pipeline using memory hierarchy optimization or ISP/NSP. Fig.~\ref{fig:motiv}(a) shows where each IVF stage executes and what data are transferred outside storage, while Fig.~\ref{fig:motiv}(b) quantifies both overall data movement and re-ranking-stage traffic, which typically dominates and is hardest to process in storage due to its high accuracy requirement. Across these designs, intermediate data are still exported for off-storage processing, leaving data movement as the key bottleneck.
For example, SPANN~\cite{SPANN} adopts a DRAM--SSD hybrid design that keeps centroids in host DRAM and stores large inverted lists on SSD. FusionANNS~\cite{fusionanns} further extends this design to a CPU/GPU/SSD hierarchy, incorporating PQ-based filtering and raw-vector re-ranking from SSD. In both cases, later-stage candidate processing remains on the host side, requiring intermediate data to move between storage and host memory.
RAGX~\cite{10.1145/3695053.3731032} and related near-storage and PIM-based designs~\cite{smartssd,UPMEM,WuATC25PIMANN,ChenSC25DRIMANN,ChenArxiv24MemANNS} push portions of computation closer to storage, reducing host-side traffic. However, they still rely on off-array processing and require intermediate data to be transferred to near-storage processors across multiple stages of the pipeline, leaving data movement as the dominant bottleneck.
REIS~\cite{reis} moves cluster selection and coarse binary search into NAND, but still performs fine-grained INT8 re-ranking on the controller. However, due to the limited precision of binary screening, REIS requires both a large number of clusters (e.g., high $N_{nlist}$) and a large re-ranking candidate set to maintain accuracy. Consequently, large intermediate data must be fetched into the SSD controller for final re-ranking, incurring high latency from TLC/QLC reads~\cite{takai2019analysis}.


These limitations motivate D-NOVA, which performs 
entire IVF three stages in storage using DTS-based in-array similarity evaluation (Section~\ref{sec:ddbam}) and in-array accumulation, ensuring that only compact metadata—rather than raw embeddings or large intermediate candidate—leaves the array. By doing so, Fig.~\ref{fig:motiv}(b) shows that D-NOVA achieves the highest overall and re-ranking stage data movement reduction compared to prior designs.

\vspace{-0.1cm}

\subsection{Overview of 3D NAND flash and Challenges} \label{sec:dbam}

\begin{figure}[t]
    \centering
    \includegraphics[width=0.98\linewidth]{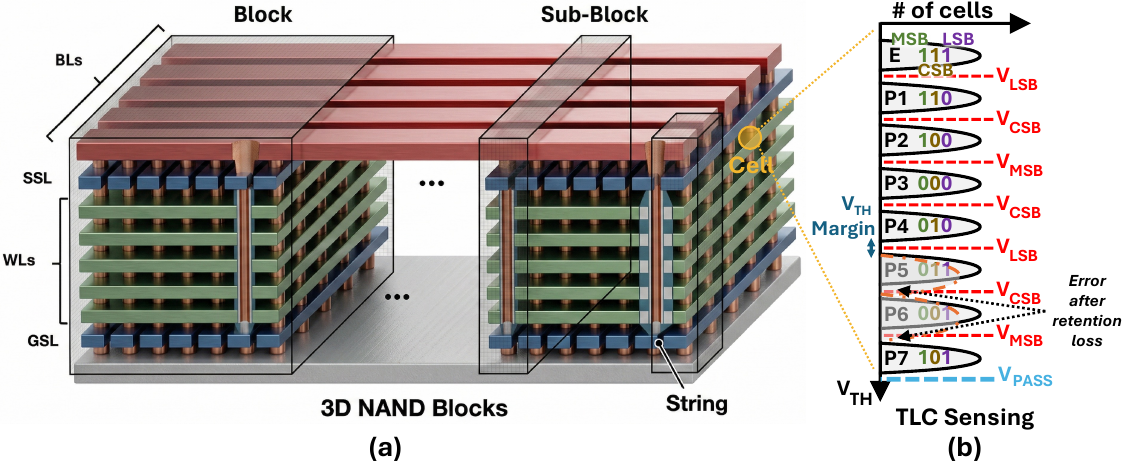}
    \caption{(a) 3D NAND block structure, and (b) $V_{\mathrm{TH}}$ distribution of NAND TLC.}
    \label{fig:fenand_dbam}
\end{figure}

Fig.~\ref{fig:fenand_dbam}(a) shows a 3D NAND block consisting of vertical strings intersected by horizontal wordlines (WLs). A transistor cell is formed at each WL–string intersection, and the cells within a string are connected in series between a bitline (BL) at the top and a ground select line (GSL) at the bottom. Each string is controlled by a string select line (SSL) and a GSL, enabling selective activation during read/program operations. Multiple strings are grouped into sub-blocks, where BLs connect them to peripheral sense amplifiers (SAs) and page buffers, which determine the page-level read granularity.

During a read operation, all BLs in one SSL group (strings sharing the same SSL) are sensed in parallel while one WL is enabled at a time. The selected WL is driven by a read voltage ($V_{\mathrm{read}}$), while unselected WLs are biased with a pass voltage ($V_{\mathrm{pass}}$) to ensure they remain conductive. If $V_{\mathrm{read}}$ exceeds the selected cell's threshold voltage ($V_{\mathrm{TH}}$), the cell conducts, and the string connects to the BL; otherwise, the string remains disconnected.
In modern multi-level 3D NAND, TLC (3~bits/cell) and QLC (4~bits/cell) encode $2^b$ $V_{\mathrm{TH}}$ states per transistor ($b{=}3$ or $4$), increasing density but requiring multiple sensing steps to determine the stored value. 
Conceptually, a $b$-bit cell requires $2^b{-}1$ $V_{\mathrm{read}}$s (Fig.~\ref{fig:fenand_dbam}(b)), applied sequentially during readout. In practice, controllers employ multi-level sensing and read--retry schemes to reduce the average sensing cost per read~\cite{cai2017error}. Even with these optimizations, QLC readout remains several times slower and more energy-intensive than an SLC sense.

One WL layer in a single SSL group (i.e., a sub-block), along with its associated BLs, constitutes a logical page that can be read in a single sensing cycle (e.g., 2~KB page with 16K BLs). Since all data in a page are accessed simultaneously, each vector is mapped entirely within a single page by distributing its embedding dimensions across BLs in the same SSL-selected string group. 
%
Unlike ReRAM- and PCM-based crossbar arrays, which implement analog PIM via current accumulation from multiple cells connected in parallel to each BL~\cite{song2025hybrid, fan2024specpcm}, NAND flash strings are serially connected and operate in a digital manner, where each cell contributes a binary conduction state (i.e., connected or disconnected).
To accommodate this structural and computational difference, D-NOVA introduces a new distance metric, \textit{DTS}, that enables accurate in-storage search directly within 3D NAND arrays with negligible modification to the existing structure, as described in Section~\ref{sec:ddbam}.

\vspace{-0.2cm}

\section{D-NOVA}

This section presents the overall D-NOVA architecture and explains how the proposed in-storage DTS retrieval operates across the plane.
D-NOVA combines NAND-string threshold sensing, page-buffer/peripheral deficit accumulation, and compact score/metadata movement to execute the three-stage IVF retrieval flow inside the storage device.
We first describe the SSD-level organization---including the controller, channels, dies, planes, and 3D NAND blocks---that enables in-storage computation across NAND arrays and the controller. We then detail the DTS mechanism and its overhead, and outline how it integrates with the IVF-based three-stage retrieval flow.

\subsection{Overall Architecture}

\begin{figure}[t]
\centering
\includegraphics[width=1\columnwidth]{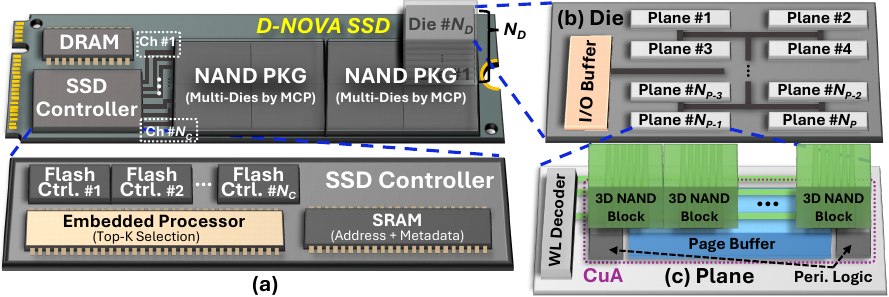}
\caption{(a) Overall architecture of D-NOVA SSD connected with SSD controller, 
(b) die with multi-planes, and 
(c) plane containing 3D NAND blocks, and peripheral circuitry.} 
\label{fig:D-NOVA_overview}
\end{figure}

D-NOVA is implemented as an in-storage accelerator fully integrated within an SSD system, as shown in Fig.~\ref{fig:D-NOVA_overview}. The system consists of an SSD controller and a D-NOVA storage chip. The SSD controller hosts flash controllers (one per channel) and an embedded processor connected to SRAM, which performs lightweight sorting operations such as quicksort~\cite{reis, hoare1962quicksort, singh2017novel} or quickselect~\cite{reis, mahmoud1995analysis, martinez2001optimal} on intermediate results from each retrieval stage, as well as metadata management for cluster centroids and Top-$K$ candidate lists.
Each incoming query is broadcast from the controller to all active channels for parallel retrieval.
The D-NOVA storage system comprises $N_C$ independent channels, each connected to $N_D$ dies within a single MCP (multi-chip package)~\cite{na20211, kim20157}.
The channel-level parallelism enables concurrent execution of multiple queries, while each die performs independent in-storage searching. 
Inside a die, $N_P$ planes are organized via H-Tree bus~\cite{xu2023proxima}, enabling balanced signal distribution and low-latency communication across planes. 
Each plane includes an array of 3D NAND blocks, which serve as the fundamental computational storage units.
Each NAND block consists of 16K BLs, 128 WLs, and 16 SSLs, and multiple NAND blocks within a plane share the same BL array, as shown in Fig.~\ref{fig:fenand_dbam}(a).
Each BL is connected to a shared SA and a local page buffer that temporarily stores binary sensed outcomes. 

Both embedding vectors and raw document chunks reside in QLC, with vectors stored in a dedicated embedding region where each cell encodes an INT4 value, and document chunks stored in a separate document region. Each embedding maintains a compact pointer to its corresponding document. After the Top-$K$ embeddings are selected, their identifiers are translated locally into document addresses, and the corresponding text payloads are fetched directly from the document region. This unified in-storage organization enables D-NOVA to perform vector retrieval entirely within the SSD, transferring only the resulting documents to the host.




\subsection{Proposed Complement-Based DTS Scheme} \label{sec:ddbam}

For accurate in-storage similarity evaluation on dense INT4 embeddings, D-NOVA introduces a complement-based \emph{Dual-Bound Tight Similarity Sensing (DTS)} scheme tailored for the NAND string. 
DTS narrows the candidates by applying upper and lower bounds (\textit{UBS} and \textit{Comp UBS}) without modifying the array structure.
While simple bound-based search has been applied to highly error-tolerant workloads such as mass spectrometry using hyperdimensional computing~\cite{pinge2025fenoms}, this naive strategy does not work in accuracy-sensitive RAG, motivating the proposed DTS design.





\begin{figure}[t]
    \centering
    \includegraphics[width=1\linewidth]{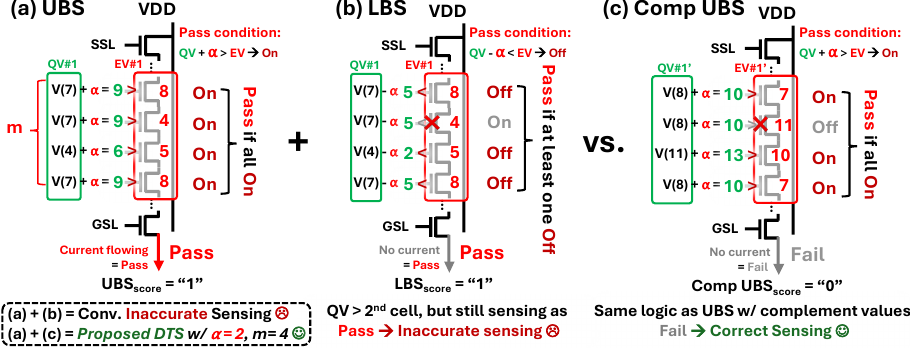}
    \caption{Overview of (a) UBS, (b) LBS, and (c) Comp UBS.}
    \label{fig:ddbam_overview}
\end{figure}

In the D-NOVA configuration, unlike conventional NAND mappings where vectors are stored within a single page along the WL direction, each embedding vector is mapped along a vertical string, with its dimensions stored in NAND cells across successive WL layers, as shown in Fig.~\ref{fig:ddbam_overview}. Since these embeddings are not conventionally read as QLC pages during DTS scoring, D-NOVA maps INT4 values to sequentially ordered $V_{\mathrm{TH}}$ levels, rather than Gray-coded logical page labels. In contrast, raw document chunks remain stored in the conventional page-wise layout to enable fast fetching after Top-$K$ retrieval, and can use standard Gray-coded QLC encoding for normal NAND reads.
During search, the query vector (QV) is given as WL voltages, each of which is proportional to the magnitude of  QV's corresponding element.
But, unlike the conventional read operation, multiple ($m$=1 - 8) WLs are enabled at a time. 3D NAND devices already support multi-WL activation modes~\cite{park2022flash, Lee2015,Sharon2014, jp2003222422a, salehi2022memory} with a negligible area and latency overhead. The details of the \(m\) configuration and the associated overheads are discussed in 
Section~\ref{sec:m_adaptive} and  Section~\ref{sec:overhead}, respectively.
As an example in Fig.~\ref{fig:ddbam_overview}, we assume $m{=}4$ and consider four consecutive dimensions of an embedding vector (EV) [8, 4, 5, 8] and a QV [7, 7, 4, 7], which are stored as INT4 values, with the EVs encoded in cell $V_{\mathrm{TH}}$ levels.
These four dimensions are sensed together in one cycle by enabling one SSL and four WLs, allowing the SA to detect the combined binary conduction state of the string (on/off). This fully digital sensing ensures robust operation compared to analog PIMs; Noise effects from process variation and the impact of the potential increase in on-resistance due to multi-WL enabling are discussed in Section~\ref{alphasensitivity} and Section~\ref{sec:overhead}, respectively.

\noindent\textbf{Upper-Bound Similarity Sensing (UBS):}
During UBS, each WL is driven with a voltage slightly higher than the corresponding query vector (QV) element, i.e., $V_{\mathrm{WL}} = q_i + \alpha$, where $\alpha$ is a tolerance margin (set to $\alpha{=}2$ in Fig.~\ref{fig:ddbam_overview}(a)) and $q_i$ denotes the $i$-th element of the QV.
If the stored $i$-th embedding element ($r_i$) satisfies $r_i < q_i + \alpha$, the applied WL voltage exceeds the cell $V_{\mathrm{TH}}$, turning on the conduction path. When all $m$ cells in a group satisfy this condition (Fig.~\ref{fig:ddbam_overview}(a)), the string conducts, and the group is counted as a pass.
This mechanism ensures that \textit{all} elements of the stored vector remain below an upper bound defined by the query, i.e., $q_i + \alpha$. Formally, the upper-bound score ($\mathrm{UBS_{score}}$) is defined as: 
\begin{equation} \mathrm{UBS_{score}}_j=\prod_{i=mj}^{mj+m-1}[\,r_i< q_i+\alpha\,]. \end{equation}
where $j$ denotes the index of the $m$-cell group sensed in parallel. UBS thus enforces a strict all-or-nothing criterion, as all cells in the group must satisfy the condition for the group to pass. 



\noindent\textbf{Lower-Bound Similarity Sensing (LBS):} As UBS checks only the upper bound, we employ LBS to complete the dual-bound check.
LBS applies $V_{\mathrm{WL}} = q_i - \alpha$ to check whether $r_i > q_i - \alpha$. Since the applied voltage is slightly lower than the corresponding QV element, it falls below the threshold of matching cells, turning off the conduction path.
Due to the serial nature of the NAND string, if \textit{any} of the $m$ cells satisfies this condition, the string becomes non-conductive, and the group is marked as a pass ($\mathrm{LBS_{score}}=1$).
Formally, the lower-bound score ($\mathrm{LBS_{score}}$) is defined as:
\begin{equation} \label{func:LBS} \mathrm{LBS_{score}}_j=1-\prod_{i=mj}^{mj+m-1}[\,r_i<q_i-\alpha\,]. \end{equation}
Because the serial NAND string reports a “1” when even a single element satisfies the bound, LBS is inherently much looser than UBS, as illustrated in Fig.~\ref{fig:ddbam_overview}(b). This lenient behavior is tolerable in noise-resilient workloads such as hyperdimensional computing~\cite{pinge2025fenoms, yang2024fslhdnn}, where extremely high-dimensional vectors ($>$10K dimensions) can statistically absorb sensing errors, but it introduces false positives in dense embeddings, degrading recall by over 10\% compared to cosine or L2 search, thereby motivating a new distance metric.

\noindent\textbf{Complement Upper-Bound Similarity Sensing (Comp UBS):}
To preserve strictness while extending coverage, D-NOVA replaces the relaxed lower-bound path (LBS) with a complement-based sensing mechanism. 
As shown in Fig.~\ref{fig:ddbam_overview}(c), each INT4 value $r_i$ in the embedding vector is stored with its 4~bits complement $r'_i{=}15{-}r_i$ (and similarly $q'_i{=}15{-}q_i$ for the query). This complement transformation reverses the inequality, enabling the use of UBS in place of LBS.
Formally, the Comp UBS is written as: 
\begin{equation}
\mathrm{Comp~UBS_{score}}_j = \prod_{i=mj}^{mj+m-1} [\,r_i' < q_i' + \alpha\,],
\end{equation}
This complement sensing eliminates false positives that LBS would otherwise produce, restoring fine-grained discrimination across INT4 embeddings.
The data arrangement of normal and complementary embeddings and their potential data volume overhead are discussed at the end of this section.

\noindent\textbf{Final Scoring with both metrics:}
The final dual-bound score is obtained by aggregating UBS and Comp UBS results across all the $m$-cell groups as follows:
\begin{equation}
\mathrm{Score}_{\mathrm{DTS}} = \sum_{j=1}^{D/m} (\mathrm{UBS_{score}}_j + \mathrm{Comp~UBS_{score}}_j) = \sum_{j=1}^{D/m} \mathrm{DTS}_j,
\end{equation}
where $D$ is the embedding dimension and $m$ is the number of WLs sensed in parallel.
A higher $\mathrm{Score}_{\mathrm{DTS}}$ indicates that more groups fall within valid bounds in both the original and complement domains, providing finer discrimination for in-array matching.
Beyond the search capability itself, aggregation across multiple $m$-cell groups serves as an important mechanism for providing robustness and resilience. Even if individual groups fail due to process variation, their impact is mitigated through accumulation, as errors are statistically averaged out across groups. This aggregation preserves the correctness of the search results despite device-level variability.



\begin{figure}[t]
    \centering
    \includegraphics[width=1\linewidth]{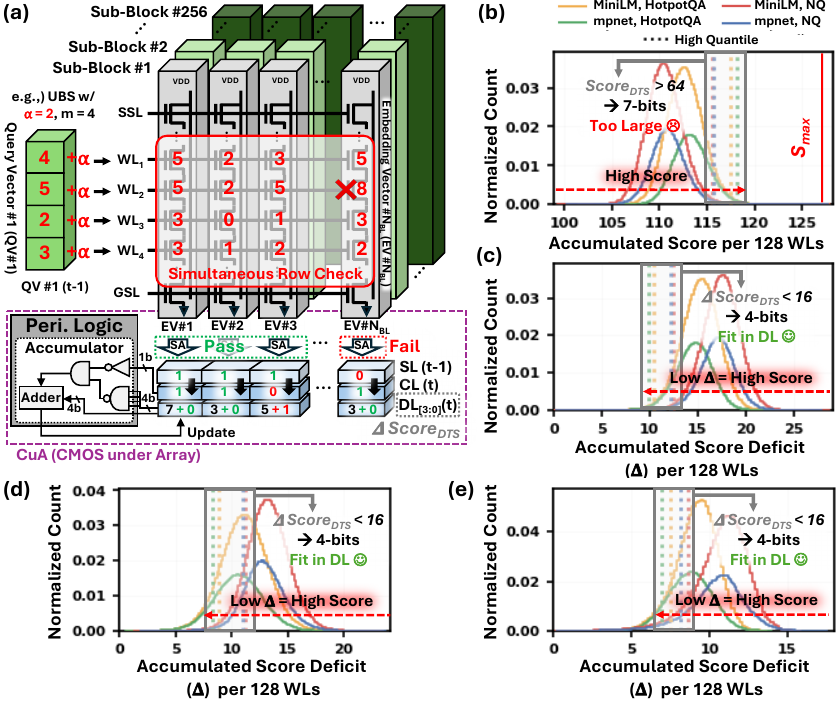}
    \caption{(a) DTS in NAND and $\Delta\mathrm{Score}_{\mathrm{DTS}}$ accumulation, (b) second stage accumulated score distributions, (c) corresponding deficit ($S_{\max}-S$) distributions,
    and (d) first, and (e) third stage deficit distributions under $m{=}(1,1,1)$.}
    \label{fig:nand_ISP}
\end{figure}

\begin{figure*}[t]
    \centering
    \includegraphics[width=0.9\linewidth]{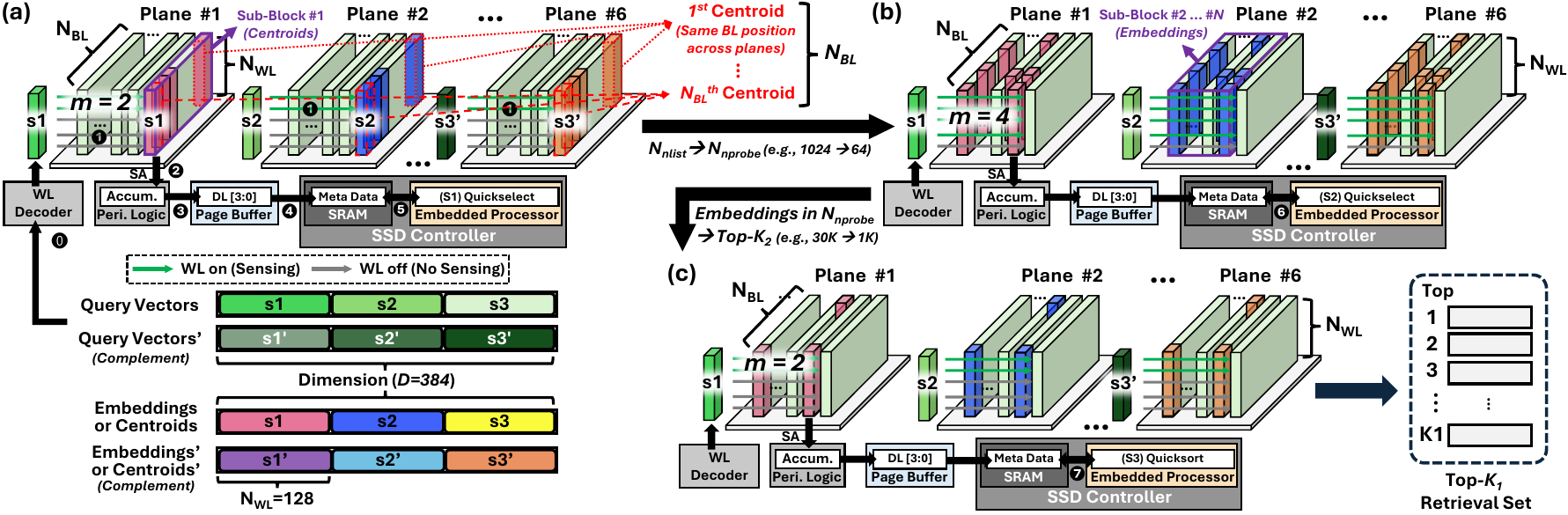}
    \caption{Vector mapping and IVF data flow in D-NOVA for the (a) first, (b) second, and (c) third stages.} 
    \label{fig:ivf_mapping}
\end{figure*}

\noindent\textbf{Delta Scoring for Compact Accumulation:}
In this design, the scores are accumulated across all the $m$-cell groups, requiring wide per-BL storage to hold the accumulated values. In the worst case ($m=1$), accumulation over 128-WLs NAND string yields a maximum raw score of $S_{\max}=128$, requiring at least 7 bits per BL for brute-force accumulation.
Fig.~\ref{fig:nand_ISP}(b) shows the second stage score distributions across datasets and encoders, where accumulated values range from $\sim$105 to 120, requiring 7-bit representation. Such wide storage cannot be embedded within the page buffer and does not fit within its storage budget, necessitating additional storage or off-array aggregation. 
To address this challenge, D-NOVA instead accumulates the \emph{score deficit} from the window-level maximum ($S_{\max}$), defined as $\Delta\mathrm{Score}_{\mathrm{DTS}} = S_{\max} - S$, where $S$ is the accumulated score over the  128-WLs string.
As shown in Fig.~\ref{fig:nand_ISP}(c), the deficit values remain small, ranging from 5 to 13 even at high quantiles, indicating that $\Delta\mathrm{Score}_{\mathrm{DTS}}$ can be represented within 4 bits. Fig.~\ref{fig:nand_ISP} (d) and (e) show that the fine-grained first and third stages also stay within the 4-bit range, with high quantiles of 7--12 and 7--9 under conservative $m=(1,1,1)$. Practical larger-$m$ settings further reduce the accumulated deficit range, increasing the margin for 4-bit storage.

We therefore reuse the existing 4-bit data latch (DL) for QLC in the page buffer to store $\Delta\mathrm{Score}_{\mathrm{DTS}}$, as shown in Fig.~\ref{fig:nand_ISP}(a), eliminating the need for additional wide per-BL storage. For the score deficit calculation, each 1-bit DTS outcome for an $m$-cell group is first captured in the existing sensing latch (SL) and then transferred to the cache latch (CL) in the page buffer.
A lightweight peripheral adder then updates the DL by accumulating the inverted CL value, i.e., the logical complement of the 1-bit sensing outcome. Thus, a passing group contributes "0" to the deficit, and a failing group contributes "1". On the other hand, a 4-input NAND gate disables further updates once the 4-bit deficit reaches 1111 to avoid overflow and safely ignore low-score candidates, which does not affect the final recall rate.
After all WLs are processed, the compact delta scores are forwarded to the controller's embedded processor for final ranking and aggregation. As the lightweight accumulators are shared across multiple BLs in a time-multiplexed manner, it incurs only minimal overhead, as discussed in Section~\ref{sec:overhead}.

\noindent\textbf{DTS Complement Data Overhead and Speed Benefit:}
DTS stores paired 4-bit values (EV and EV$'$) in different planes, preserving the same data volume as an INT8-based system NSP baseline (described in Section~\ref{sec:baseline_eval}) and REIS~\cite{reis} while enabling parallel search across planes. Unlike REIS, which allocates separate SLC regions for binary vectors—thereby wasting three bits per QLC cell—and additional INT8 data for re-ranking, D-NOVA maintains a uniform 4-bit format aligned with QLC, eliminating redundant capacity overhead. In terms of latency, DTS replaces multi-level QLC reads with a small, fixed number of binary sense operations, avoiding the 7--15 iterative reads by sweeping $V_{read}$, which is required for full QLC decoding shown in Fig.~\ref{fig:fenand_dbam}(b).
The total sensing cycles ($T$) are given by $T_{\text{DTS}} = (D/m)\times 2$ for DTS and $T_{\text{QLC}} = D \times N_{\text{rd}}$ for conventional QLC reads,
where the factor of 2 accounts for UBS and Comp UBS, and $N_{\text{rd}}$ is the number of sense operations per multi-level cell ($2^b{-}1$, e.g., 7 for TLC and 15 for QLC).
Consequently, DTS achieves up to a 9.3× reduction in sensing cycles and comparable speedup over REIS without additional storage overhead.

\subsection{Three-Stage IVF Retrieval and Data Flow} \label{sec:data_flow}

D-NOVA implements an end-to-end in-storage IVF retrieval hierarchy, storing all centroids, embeddings, and document chunks directly within the SSD. Unlike prior ISP/NSP architectures such as REIS~\cite{reis} and RAGX~\cite{10.1145/3695053.3731032}, it executes the entire three-stage similarity search within the array with minimal data transfer to the embedded processor. 
Intermediate similarity results are accumulated in plane-level peripherals, while only final scores and compact metadata (e.g., addresses and centroid identifiers), which are much smaller than raw embeddings, are stored in the controller memory, minimizing data movement.

Fig.~\ref{fig:ivf_mapping} shows the vector mapping and retrieval data flow for each stage.
For exposition purposes, assume each vector is represented as a 384-$D$ INT4 vector along with its 4-bit complement for DTS computation, divided into six 128-$D$ segments to fit within 128 WLs. In the Fig.~\ref{fig:ivf_mapping}(a), the six segments—original (s1, s2, and s3) and complement (s1', s2', and s3')—are each vertically mapped onto separate NAND strings located in different planes, but at the same sub-block position in their planes. These six strings together form a logical group that is sensed simultaneously within a single cycle.

During the \textbf{first stage}, for example, centroid data is deliberately distributed across the first sub-block of each plane, and multiple centroids are packed to fill the first sub-block of each plane as shown in Fig.~\ref{fig:ivf_mapping}(a).
The first sub-block stores up to 16K centroids (considering 16K BLs), and the rest store embeddings or document chunks.
During DTS, \blackcircled{0}the query vector is broadcast from the SSD controller to all active channels for parallel processing. 
Each channel delivers it to multiple dies and planes via the H-Tree routing network~\cite{xu2023proxima, jouppi2021ten, song2025hybrid}.
\blackcircled{1}The query is applied as WL voltages through the WL driver for DTS-based centroid search (Fig.~\ref{fig:ivf_mapping}(a)). We use small $m$ values (e.g., 1-4) for this stage to maintain high precision centroid matching (the $m$-adaptive sensing mechanism is detailed in Section~\ref{sec:m_adaptive}).
%
\blackcircled{2} Each segment-level 1-bit outcome is captured by SA and stored in the SL and transferred to the CL, and then \blackcircled{3} the $\Delta\mathrm{Score}_{\mathrm{DTS}}$ is incrementally accumulated in the page buffer DL via a lightweight peripheral accumulator.
Because the centroid segments reside in different planes and each sub-block holds multiple centroids' segment, D-NOVA performs comparisons over a large volume of centroids in parallel both across planes and within each sub-block.
After score accumulation, \blackcircled{4}the 4-bit accumulated scores from each plane are merged at the SSD controller to form a 6-bit $\Delta\mathrm{Score}_{\mathrm{DTS}}$ per centroid, and only lightweight metadata is temporarily stored in the controller’s on-chip SRAM. Each piece of metadata consists of a $\log_2(N_{\text{list}})$-bit centroid identifier (CID), a 6-bit distance value (DIST), and a compact address field (CADR) including the physical plane, block, SSL, die, and channel indices ($\sim$37-bit in total). \blackcircled{5}These \{CID, DIST, CADR\} entries are then transmitted to the embedded processor for lightweight quickselect reduction~\cite{reis, mahmoud1995analysis, martinez2001optimal} to locate the top $N_p$ clusters. With $N_p = 64$, the total SRAM footprint remains $< 0.5$~KB, which is negligible compared to the controller's capacity.


The \textbf{second stage} (Fig.~\ref{fig:ivf_mapping}(b)) refines retrieval by evaluating only the vectors within the chosen clusters to produce a coarse Top-$K_2$ candidate list.
Centroid identifiers (CIDs) are translated into physical NAND addresses using a small lookup table (cluster directory). 
Each directory entry maps CID and CADR to its physical coordinates—plane number, sub-block ID, and BL segment group—allowing direct activation of those planes.
This operation remains compatible with the conventional NAND access pattern, as only the addressed sub-blocks are enabled while all other structural and timing behaviors follow standard NAND operation.  
Each activated sub-block performs DTS-based comparison using larger $m$ values (e.g., 4–8) to increase throughput, as the coarse retrieval stage accounts for over 95\% of all similarity search operations.
Partial scores from each segment group are accumulated locally by their corresponding accumulators and merged in the controller, and the embedded processor identifies the Top-$K_2$ candidates using the same \blackcircled{6}quickselect procedure.
Initially, each candidate entry is represented in a \{EID (embedding ID), DIST, EADR (embedding address)\} format.
After candidate selection, to minimize memory usage, the processor retains only the \{EID, EADR\} pair of the Top-$K_2$ candidates. 
Even for a 1B-vector DB with 768-$D$ embeddings, the transient controller-side SRAM required to receive, aggregate, and buffer compact candidate metadata during Top-$K_2$ selection remains within a few megabytes, well within the available memory budget ($\sim$8~MB)\cite{arm2016cortexr8}.
In the \textbf{third stage} (precise re-ranking, Fig.~\ref{fig:ivf_mapping}(c)), 
the corresponding small set of planes and sub-blocks are reactivated for fine-grained re-ranking for higher precision, using a smaller $m$ (e.g., 1–4) to increase sensing precision. 
After accumulation, \blackcircled{7}the embedded processor performs a final quicksort~\cite{hoare1962quicksort, singh2017novel} to rank the Top-$K_1$ results, which contributes $<0.1\%$ of the total latency and energy due to the highly limited candidate set.



This data flow avoids reading full embeddings out of NAND for conventional distance computation. D-NOVA does not implement cosine similarity or L2 distance inside the NAND array; instead, each IVF stage uses DTS as a threshold-based retrieval score. The NAND cell array performs query-dependent bound-check sensing over ordered INT4 $V_{\mathrm{TH}}$ levels, page-buffer/peripheral logic accumulates compact deficit scores, and the embedded controller performs comparison-based Top-$K$ selection over compact scores and metadata. Unlike controller- or ASIC-based distance computation, which must first read QLC/TLC embeddings through multiple sensing cycles and buffer them before arithmetic scoring (e.g., Hamming-distance or dot-product scoring), D-NOVA's binary DTS sensing evaluates the threshold score directly in the storage path while preserving high-density QLC storage.


\vspace{-0.4cm}

\section{Throughput and Accuracy Enhancement}\label{adapter_section}
To further improve D-NOVA, we introduce three complementary techniques: stage-aware $m$-adaptive sensing for higher parallelism, a DTS-optimized adapter to improve recall, and a locality-aware mapping strategy for efficient vector placement.




\subsection{Stage-Aware \texorpdfstring{$m$}{m}-Adaptive Sensing}

\begin{figure}[t]
    \centering
    \includegraphics[width=0.9\linewidth]{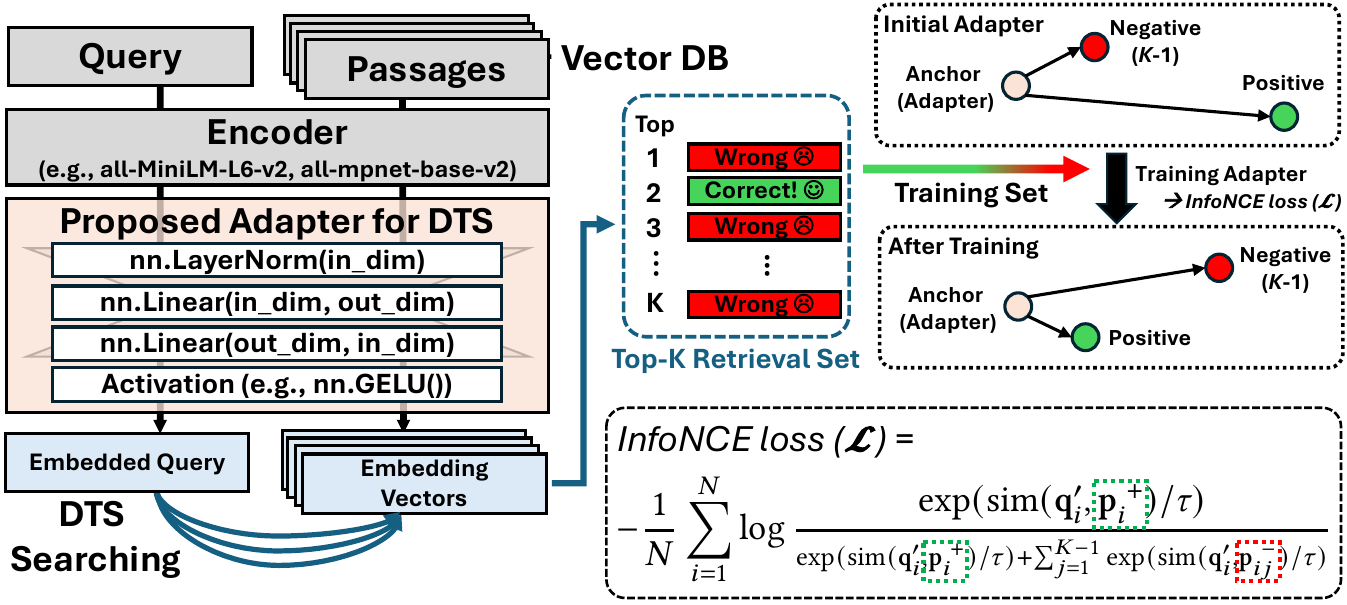}
    \caption{Adapter architecture and retrieval-aware training.}
    \label{fig:adapter}
\end{figure}

\label{sec:m_adaptive}
DTS exposes an architectural knob for intra-string parallelism via the multi-WL activation mechanism~\cite{park2022flash, Lee2015} described in Section~\ref{sec:ddbam}.
With $m$ WLs sensed in parallel within a string under a single SSL activation, where each WL corresponds to one vector dimension, each sensing step produces a single pass/fail bit for $m$ dimensions, improving per-cycle throughput.
However, increasing $m$ yields more coarse-grained decisions, since sensing produces only a group-level pass/fail outcome rather than element-wise results, reducing sensing accuracy. In contrast, smaller $m$ values preserve fine-grained decisions, enabling more accurate sensing at the cost of lower throughput, reflecting the trade-off between accuracy and efficiency.

In practice, however, not all stages of the IVF are equally sensitive to this accuracy trade-off.  
Centroid selection and Top-$K_1$ re-ranking directly determine which clusters are explored and how the final results are ordered, respectively; errors in these stages can irreversibly exclude relevant documents.  
In contrast, coarse Top-$K_2$ retrieval primarily prunes a large candidate pool into a manageable working set (typically $K_2 \gg K_1$) and can tolerate modest sensing noise. 
Accordingly, D-NOVA adopts a stage-aware configuration, where $m_s$ denotes the parallel sensing degree at stage $s$: $m_1$ and $m_3$ are kept small for the accuracy-critical centroid selection and final re-ranking stages (e.g., $1$--$4$), while $m_2$ is increased during coarse Top-$K_2$ retrieval (e.g., $4$--$8$), where throughput dominates and modest sensing noise can be tolerated. This stage-wise rule provides the design guideline for selecting $m$.
Section~\ref{sec:result_m} shows this asymmetric configuration preserves accuracy, while Section~\ref{sec:system_results} shows that it delivers 4.3--12.1$\times$  higher throughput than REIS; Sections~\ref{alphasensitivity} and~\ref{sec:overhead} further confirm robust operation under noise with minimal hardware overhead.



\subsection{Contrastive Adapter Co-Design}\label{adapter_section}

Existing text encoders such as all-MiniLM-L6-v2~\cite{wang2020minilm} and all-mpnet-base-v2~\cite{song2020mpnet} are optimized for continuous similarity metrics such as cosine similarity or L2 distance, whereas D-NOVA’s DTS operates in a discrete, threshold-based metric space. Applying DTS directly to these embeddings introduces a metric mismatch that can degrade recall, especially for semantically similar vectors separated by small numeric deviations. To mitigate this mismatch, we introduce a lightweight adapter applied to the query at runtime to reshape encoder outputs into a DTS-aligned representation, as shown in Fig.~\ref{fig:adapter}.
Training uses a retrieval-aware InfoNCE loss~\cite{oord2018representation, parulekar2023infonce} constructed from DTS retrieval results rather than random negatives. For a set of queries $\{q_i\}$, let $p_i^{+}$ denote the ground-truth positive passage for query $q_i$, and let $p_{ij}^{-}$ denote the $j$-th negative passage from the Top-$K_{\text{cand}}$ candidate pool retrieved by DTS search. The adapter is trained on these query-passage pairs by minimizing the InfoNCE loss ($\mathcal{L}$) shown in Fig.~\ref{fig:adapter}. Here, $q_i'$ denotes the adapter-transformed query embedding, $\tau$ is the temperature parameter, and $\mathrm{sim}(\cdot,\cdot)$ denotes cosine similarity. Unlike conventional contrastive training~\cite{hafidi2022negative}, which uses randomly sampled embeddings as negatives, our method uses hard negatives obtained from DTS itself. These DTS-mined negatives are more informative, 
because they emphasize near-boundary false positives, enabling the adapter to better align the query representation with DTS’s dual-bound metric.

Quantitatively, the adapter significantly improves recall by reducing the average Recall@100 gap to IVF from 3.77\%--4.54\% to 1.43\%--1.76\%, as detailed in Section~\ref{sec:accuracy}. Training is a one-time \textit{offline} cost during embedding generation and incurs no latency or computation overhead during retrieval. At inference, D-NOVA applies only a lightweight query-side adapter while keeping passage embeddings fixed in 3D NAND. This adaptation adds only 0.5\%--1\% overhead to query encoding time and is negligible in end-to-end LLM workloads. The adapter itself is compact (0.4M parameters for 384-$D$ embeddings) and can be efficiently trained or fine-tuned on standard GPUs, while inference can be performed on either a GPU or a CPU host during retrieval, enabling low-cost and flexible deployment. This HW/SW co-design sustains near-FP32-IVF recall while preserving D-NOVA’s zero-overhead in-storage execution. This query-side design preserves the RAG database-update path. New documents can be added through normal index insertion: encode the document, quantize it into D-NOVA's INT4/complement representation, assign it to the nearest centroid, and append it to that centroid's IVF list. This does not require retraining the LLM, rewriting existing passage embeddings, or modifying the storage-side DTS pipeline. If a large database update changes the query/workload distribution, the compact adapter may be optionally refreshed offline as part of index maintenance; ordinary document insertion does not require this, and the in-storage execution flow remains unchanged.
\vspace{-0.1cm}

\subsection{Locality-Aware Mapping}\label{lamapping}

\begin{figure}[t]
    \centering
    \includegraphics[width=0.9\linewidth]{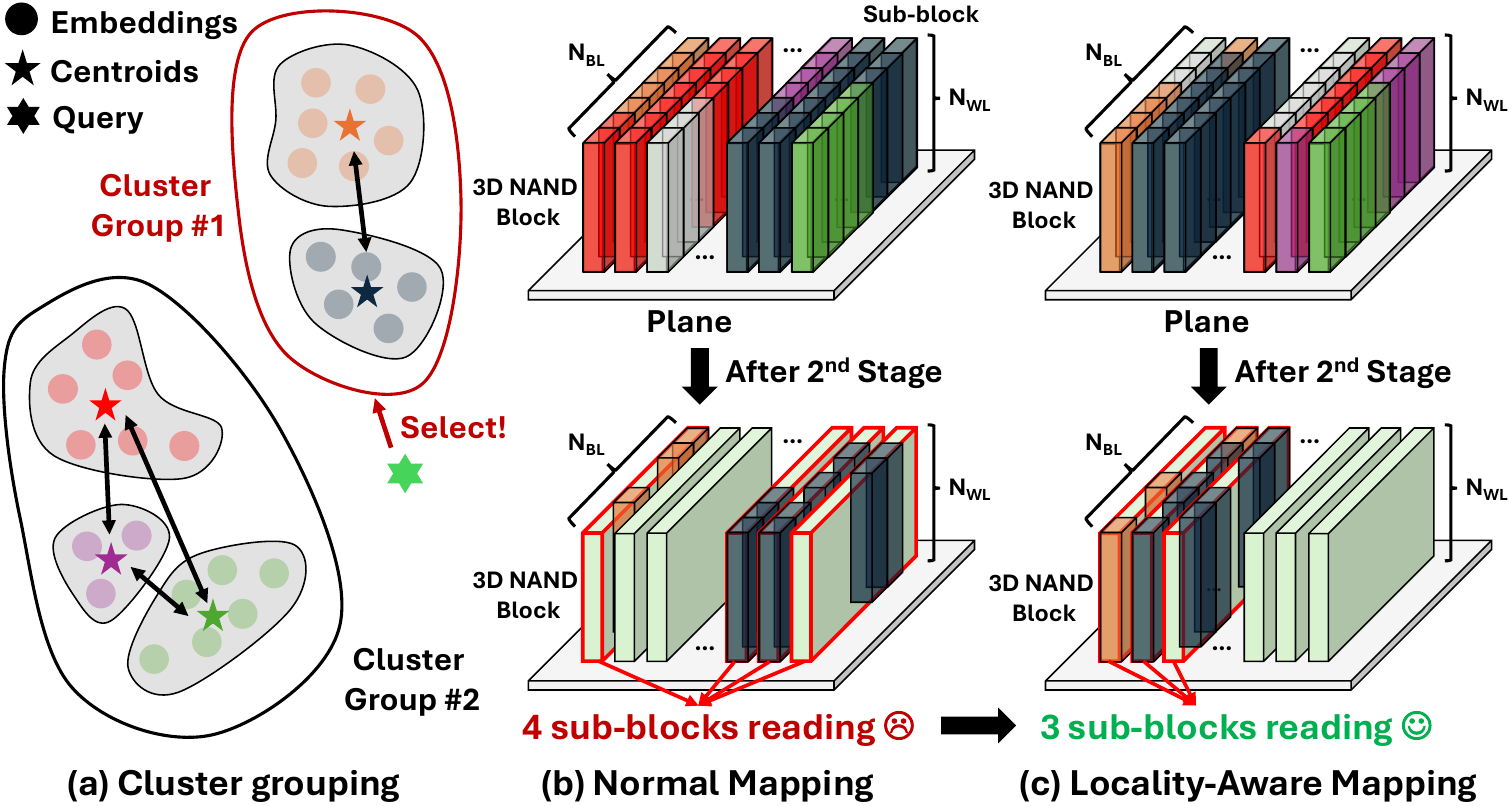}
    \caption{(a) Overview of cluster grouping, (b) normal vector mapping, and (c) proposed locality-aware mapping.} 
    \label{fig:cluster_mapping}
\end{figure}

To reduce sub-block activations per query, we propose locality-aware mapping, which co-locates relevant data by reorganizing IVF-list placement.
This is especially important in the re-ranking stage, where a small Top-$K_1$ subset of vectors may otherwise be scattered across strings, forcing many sub-block activations. 
The baseline (Fig.~\ref{fig:cluster_mapping}(b)) uses a simple, first-fit layout that packs inverted lists (the vectors assigned to each centroid) into sub-blocks by size, allocating a new sub-block when capacity is exhausted. This layout is implementation-friendly but ignores centroid geometry and cross-list query locality. In contrast, the proposed locality-aware mapping introduces a two-step procedure. First, it builds a \emph{similarity chain} over centroids using a greedy nearest-neighbor walk~\cite{nns}: starting from the largest list (by size), it repeatedly appends the unused centroid whose centroid vector has the smallest distance to the last centroid in the chain. 
As described in Fig.~\ref{fig:cluster_mapping}(a), this produces a one-dimensional ordering in which neighboring IVF lists are geometrically close and likely to be co-accessed. 
The lists are then packed into sub-blocks by traversing this chain and applying the same first-fit rule as in the baseline, naturally grouping co-accessed lists into the same or nearby sub-blocks (Fig.~\ref{fig:cluster_mapping}(c)).
With locality-aware mapping, second- and third-stage candidate accesses concentrate in fewer regions, reducing sensing operations per query and improving throughput. As quantified in Section~\ref{sec:la}, this lowers the median activation footprint by about 40\%.

\vspace{-0.1cm}


\section{Experimental Setup}
\label{sec:method}


\subsection{Benchmarks and Encoder}
\label{benchmarks}

We evaluate D-NOVA on six retrieval benchmarks representative of RAG-oriented dense retrieval: NQ (2.68M documents; open-domain QA)~\cite{nqdataset}, HotpotQA (5.23M; multi-hop QA)~\cite{hotpotdataset}, FEVER (5.42M; fact verification)~\cite{feverdataset}, and FiQA (57.6K; financial QA)~\cite{fiqadataset} from BEIR~\cite{beirdataset}, wiki\_en (1.01M; open-domain QA) from AIR-Bench~\cite{airbenchdataset}, and MIRACL-English (32.9M; ad-hoc retrieval) from MIRACL~\cite{miracl}. All methods use the same text encoders for fair comparison: all-MiniLM-L6-v2 (MiniLM, 384-$D$)~\cite{wang2020minilm} and all-mpnet-base-v2 (mpnet, 768-$D$)~\cite{song2020mpnet}.
We use FP32-IVF as the full-precision software accuracy reference: it uses the same IVF-Flat index, FP32 embeddings, and cosine similarity, without DTS quantization or NAND sensing constraints. This FP32-IVF reference corresponds to the CPU-IVF baseline whose system configuration is described in Section~\ref{sec:baseline_eval}.
For D-NOVA and the DTS variants, we attach our proposed adapter to the encoder outputs and train it once \textit{offline}, as described in Section~\ref{adapter_section}, while keeping the adapter frozen during retrieval.
Unless otherwise noted, all systems use the same IVF-Flat design with dataset-specific search parameters. 
We use $N_{\mathrm{list}}=1024$ for NQ, FiQA, and wiki\_en, $2048$ for HotpotQA and FEVER, and $8192$ for MIRACL-English. We evaluate $N_{\mathrm{probe}}=\{64,128\}$ for NQ, $\{32,64\}$ for FiQA and wiki\_en, $\{128,256\}$ for HotpotQA and FEVER, and $\{256,512\}$ for MIRACL-English.
Across all datasets, we keep $K_2=1000$ coarse candidates and $K_1=100$ final results fixed for both FP32-IVF and all DTS configurations. Therefore, the recall results reflect the DTS metric, adapter, and sensing configuration under the same candidate budgets, rather than differences in IVF search effort.


\vspace{-0.1cm}

\subsection{In-House DTS \& Adapter Simulator}
\label{simulator}



\begin{table}[h]
\centering
\caption{Simulator Configuration}
\label{tab:sim_config}
\begin{tabular}{l l}
\toprule
\textbf{Component} & \textbf{Configuration} \\
\midrule
Read Energy & 294.9 nJ/page~\cite{nandenergy} \\
Read Latency & 30 $\mu$s (SLC), 140 $\mu$s (QLC)~\cite{takai2019analysis} \\
Circuit Simulation & Spectre 19.1, PTM-MG 14nm~\cite{zhao2007predictive} \\
String/Cell Config. & 128 WLs / QLC within 8.5V window~\cite{choi2026machine}\\
\bottomrule
\end{tabular}
\end{table}

We develop a cycle-accurate, event-driven simulator to model D-NOVA’s DTS operation on a realistic 3D NAND array. Table~\ref{tab:sim_config} summarizes the simulator configuration.
The simulator tracks WL activations, page-buffer events, and controller-side reductions at cycle granularity, translating them to latency and energy using measured NAND timing parameters and per-operation energy models from prior works~\cite{takai2019analysis,nandenergy}, rescaled to our block organization using an RC-based 3D NAND model~\cite{RC}. 
This includes the movement of compact scores and metadata through page buffers and controller SRAM, local deficit updates in the peripheral logic, and sub-block activations determined by the selected IVF lists.
To ground the system-level abstraction in device-level behavior, we perform circuit-level Monte Carlo simulations that incorporate both cell-level and peripheral non-idealities, including $V_{TH}$ mismatch and interconnect parasitics. Each Spectre netlist instantiates a precharged bitline, a 128-cell NAND string, a sense amplifier, and an output buffer using PTM-MG 14nm~\cite{zhao2007predictive} models. In the scaled PTM voltage domain, query-derived WL voltages are generated as $V_{\mathrm{WL}}=0.1+(q+\alpha)\Delta V_{\mathrm{TH}}$ over a 16-level QLC window, while unselected WLs are biased with $V_{\mathrm{PASS}}=1.20$~V. The binary DTS sensing output is determined by whether the bitline crosses the 0.40~V sense trip point during the read window. $V_{\mathrm{TH}}$ variations are injected based on QLC drift characteristics from~\cite{papandreou2020open}, modeling the open-block scenario in which $V_{\mathrm{TH}}$ distributions exhibit significant broadening and shifts. This captures sensing uncertainty from process variation and retention/drift-like threshold broadening rather than assuming ideal threshold levels; the same variation model is used throughout the $\alpha$ sweep in Sec.~\ref{alphasensitivity}, so the reported sensing accuracy and Recall@100 reflect perturbed QLC thresholds.
To account for non-idealities introduced by multi-WL enabling, the effective string resistance is scaled with the number of simultaneously activated cells ($m$), which reduces the sensing resolution at larger $m$. 
In DTS, the tolerance $\alpha$ is added to the query WL voltage used in bound checks: a larger $\alpha$ broadens the acceptance region, while smaller values enforce stricter matches.  All experiments sweep $\alpha$ under noisy $V_{TH}$ conditions to reflect the process variation and report accuracy with perturbed $V_{TH}$ from limited retention. The simulator also models $m$-adaptive grouping and locality-aware mapping, and incorporates the functionality of one-time adapter training via an integrated end-to-end training loop driven by DTS retrieval results.

\vspace{-0.1cm}

\subsection{D-NOVA Configuration} \label{sec:dnova_config}

\begin{table}[h]
\centering
\caption{D-NOVA Configuration}
\label{tab:dnova_config}
\begin{tabular}{l l}
\toprule
\textbf{Component} & \textbf{Configuration} \\
\midrule
Host Interface & PCIe 4.0 $\times$4 (7.88 GB/s)~\cite{Bae2021A1P} \\
Controller & Embedded CPU~\cite{arm2016cortexr8}, $\sim$8 MB SRAM \\
Flash Channels & 8 channels (2.0 GB/s/ch, 16 GB/s total)~\cite{9366054} \\
NAND PKG & 8 dies/ch, 12 planes/die, 256 blocks/plane \\
Block Spec & 128 WL, 16K BL, 16 SSL \\
Page Buffer & 2 KB (SLC), 8 KB latch capacity (QLC)~\cite{khakifirooz202130, leong2008random} \\
\bottomrule
\end{tabular}
\end{table}


\begin{figure*}[t]
    \centering
    \setlength{\fboxrule}{1pt}        
    \setlength{\fboxsep}{2pt}        
      \begin{minipage}{\textwidth}
        \centering
        \includegraphics[width=\linewidth]{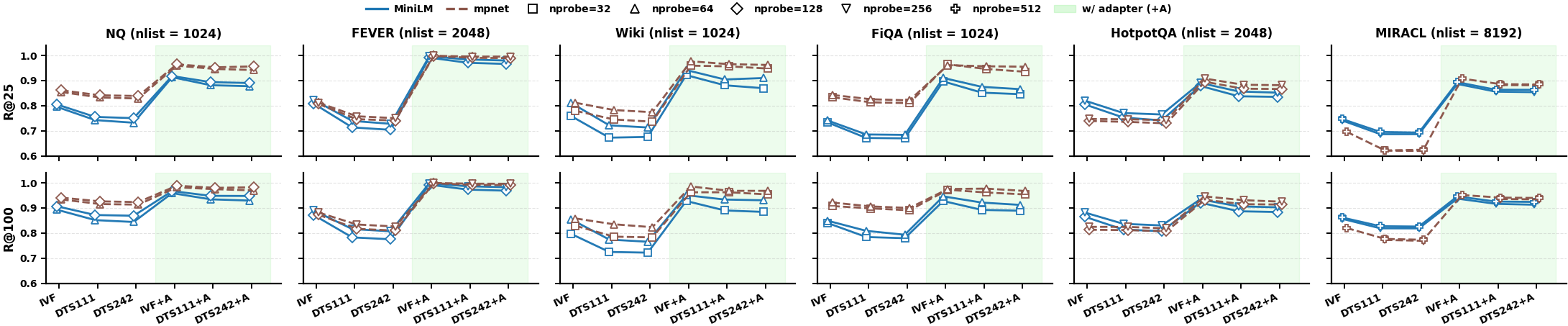}
        \caption{Recall@25 and Recall@100 for FP32-IVF, DTS111, and DTS242, each with and without the adapter, across six datasets.}
        \label{fig:recall}
      \end{minipage}
\end{figure*}

D-NOVA is implemented as a self-contained NVMe SSD subsystem. Table~\ref{tab:dnova_config} summarizes its overall system configuration.
The storage hierarchy exposes high internal bandwidth through multi-channel and multi-plane parallelism. In particular, planes within each die are interconnected via an H-Tree routing network~\cite{xu2023proxima, jouppi2021ten, song2025hybrid}, enabling balanced signal propagation and high inter-plane bandwidth ($\approx$1 TB/s), which is critical for scalable in-storage parallel search.
Although DTS relies on binary sensing by utilizing simple pass/fail comparators, D-NOVA fully preserves compatibility with standard QLC read/program operations by reusing existing page-buffer structures. 
D-NOVA employs a lightweight shared accumulator in the peripheral logic for a group of 256 BLs to aggregate CL outputs. The accumulator is implemented using digital logic and synthesized using the ASAP7 7\,nm~\cite{clark2016asap7} standard-cell library with the OpenROAD flow~\cite{ajayi2019openroad}, targeting a 1\,GHz operating frequency. The design choice and associated area overhead are discussed in Section~\ref{sec:overhead}.
The remaining on-die peripheral circuits, such as page decoders and WL/BL drivers, follow the conventional NAND configuration~\cite{reis}, preserving full compatibility with standard NAND operations (e.g., ISPP/ISPE). Consequently, by using single-cycle binary sensing instead of multi-level QLC reads, D-NOVA achieves SLC-like read latency while maintaining QLC density.

\subsection{Hardware Baselines and Metrics}
\label{sec:baseline_eval}

We evaluate D-NOVA against both a CPU-based software baseline, as they represent the software upper bound in accuracy and latency, and state-of-the-art ISP/NSP accelerators representative of current practices in large-scale vector retrieval, as described below. 

\textbf{1) CPU–IVF baseline} uses FAISS IVF-Flat on a dual-socket 128-core server~\cite{amd2023epyc9554} equipped with 1.5 TB DDR4 DRAM~\cite{micron2025ddr4sdram} and a PCIe 4.0 SSD~\cite{samsung2021pm9a3}.
All embeddings are stored in DRAM, and cosine similarity is computed in FP32 precision, providing the software upper bound unconstrained by NAND bandwidth or quantization effects.
System-level power is estimated using AMD $\mu$Prof~\cite{amduprof} for CPU components, while DRAM energy is modeled using CACTI7~\cite{balasubramonian2017cacti7}, as $\mu$Prof does not account for main-memory power. 

\textbf{2) REIS baseline}~\cite{reis} performs cluster selection and binary coarse search inside NAND, followed by INT8 re-ranking in the controller. We scale its $N_{\text{nlist}}$ and $N_{\text{nprobe}}$ parameters to align with our datasets. Binary search energy and latency for the REIS baseline are modeled using measured NAND SSD characteristics~\cite{takai2019analysis}, identical to those used for D-NOVA.
The re-ranking stage operates on the controller’s embedded processor~\cite{arm2016cortexr8} with LPDDR4 memory~\cite{micron2023lpddr4x}, where INT8 dot-product energy and latency are derived from~\cite{yazdanbakhsh2022sparse} and scaled to 7~nm following the methodology in~\cite{Stillmaker201774}. 

\textbf{3) NSP baseline} models a near-storage processor similar to RAGX \cite{10.1145/3695053.3731032}. It stores INT8 embeddings in QLC NAND, which are fetched via normal NAND read operations. An ASIC accelerator adjacent to the SSD controller executes the first- and second-stage INT8 cosine similarity searches, while the third stage performs only quicksort on the resulting candidates. For fair comparison, we use identical $N_{\text{nlist}}$ and $N_{\text{nprobe}}$ parameters and scaled processor and DRAM specifications as in REIS. This baseline reflects a near-data design that reduces host traffic but still requires full INT8 readouts and arithmetic computation outside the NAND array. 
For datasets whose indexes fit entirely in DRAM, DRAM-PIM~\cite{ChenSC25DRIMANN} could also be considered. However, it assumes a volatile DRAM-resident index, while D-NOVA targets NAND-resident embeddings and avoids full-embedding readout. Efficient DRAM-PIM ANN also often changes the index to IVF-PQ/LUT scoring, introducing a different accuracy/storage trade-off from our IVF-Flat reference. For larger RAG corpora, DRAM-PIM must populate or maintain a DRAM copy from SSD/NAND, whereas D-NOVA directly searches the NAND-resident index.

All results include NAND reading,  on-die reductions, locality-aware mapping effects, and controller-side Top-$K$ selection and metadata handling. Unless otherwise specified, the DTS tolerance margin is set to $\alpha{=}2$.  
In our evaluation, GPU-based baselines are not considered, as our study, consistent with prior work~\cite{reis, SPANN} in the retrieval stage, indicates that memory and I/O are bandwidth bottlenecks rather than computational throughput. 
We report Recall@$25$ \& @$100$, latency \& energy per query, and queries per second (QPS).

\begin{figure*}[t]
    \centering
    \begin{minipage}[t]{0.68\linewidth}
        \centering
        \includegraphics[width=\linewidth]
        {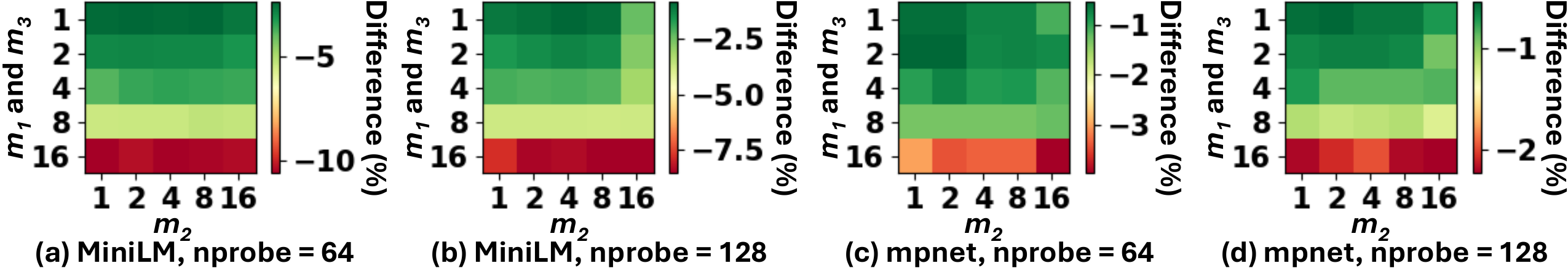} \vspace{-0.2cm}
        \caption{(a)-(d) Recall@100 degradation heatmap across different encoders for the NQ under $m$-adaptive configuration and different stage settings ($m_1, m_2, m_3$).}
        \label{fig:m_adaptive_abcd}
    \end{minipage}
    \hfill
    \begin{minipage}[t]{0.3\linewidth}
        \centering
        \includegraphics[width=\linewidth]{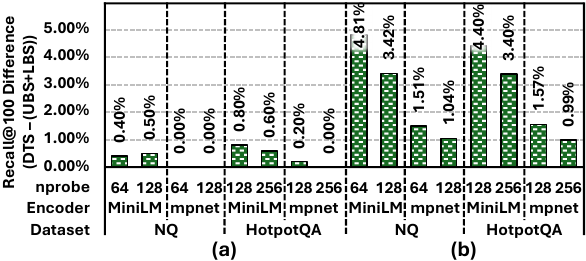} 
        \vspace{-0.2cm}
        \caption{Benefit of (a) DTS141, and (b) DTS441 vs. (UBS + LBS).}
        \label{fig:dts_benefit}
    \end{minipage}
\end{figure*}

\section{Evaluation Results}
\label{eval}

\subsection{DTS Retrieval Accuracy Results}\label{sec:accuracy}



For brevity, we denote DTS configurations as DTS$m_1m_2m_3$ (e.g., DTS242 corresponds to $m{=}(2,4,2)$). 
Fig.~\ref{fig:recall} reports Recall@25 (R@25) and Recall@100 (R@100) for FP32-IVF baseline and DTS variants under two configurations (DTS111 and DTS242), each evaluated with and without the adapter.
As shown in Fig.~\ref{fig:recall}, DTS closely tracks the FP32-IVF baseline, with only a small accuracy gap at the largest $N_{\text{nprobe}}$. The gap is consistently smaller for mpnet than for MiniLM, indicating that the denser encoder is less sensitive to the discrete DTS metric. Increasing the sensing group size $m$ from DTS111 to DTS242 introduces a modest degradation, while still maintaining stable accuracy across benchmarks and encoders.
Applying our proposed adapter significantly improves recall not only for DTS, but also for the baseline IVF, with larger gains observed for DTS as summarized in Table~\ref{tab:recall_summary}. Since the adapter is trained with DTS-mined hard negatives, it reshapes the query representation to separate relevant passages from DTS-confusing false positives.
The same trend holds across both encoders and all six benchmarks, indicating that retrieval-aware adaptation effectively compensates for the residual mismatch between the DTS metric and the FP32 retrieval space. The adapter also reduces the difference between DTS111 and DTS242, showing that part of the loss from coarser sensing can be recovered in the embedding space.
As an additional NQ database-expansion stress test, we partition the corpus into an 80\% base database and a 20\% added-document set. The adapter is trained only on the base database; the added documents are then inserted into the IVF lists, and the frozen base adapter is reused without retraining. Averaged over randomized database splits, MiniLM/mpnet, and $N_{\mathrm{probe}}\in\{64,128\}$, the frozen adapter reduces the expanded-DB DTS111-IVF R@100 gap from 4.95\% to 2.93\%, and the DTS242-IVF gap from 6.40\% to 3.04\%. This shows that newly added documents can be searched using the existing adapter while preserving the same storage-side DTS flow.


Overall, these results show that DTS preserves strong retrieval quality across diverse tasks, domains, and encoder choices, while retrieval-aware adaptation materially reduces the accuracy degradation without altering the storage-side retrieval flow.

\begin{table}[t]
\centering
\caption{Average R@100 gap across 6 datasets \& 2 encoders.}
\label{tab:recall_summary}
\vspace{2pt}
\begin{tabular}{lccc}
\toprule
\multirow{2}{*}{\textbf{Method}} 
  & \multicolumn{2}{c}{\textbf{Gap to FP32-IVF (\%)}} 
  & \multirow{2}{*}{\textbf{Adapter Gain (\%)}} \\
\cmidrule(lr){2-3}
  & \textbf{w/o Adapter} & \textbf{w/ Adapter} &  \\
\midrule
DTS111  & 3.79 & 1.45 ($\downarrow$ 2.34) & +11.48 \\
DTS242  & 4.48 & 1.75 ($\downarrow$ 2.73) & +11.87 \\
\bottomrule
\end{tabular}
\end{table}

\subsection{$m$-Adaptive Parallelism vs. Recall Accuracy} 
\label{sec:result_m}

Fig.~\ref{fig:m_adaptive_abcd}(a)–(d) show heatmaps sweeping $m_2$ along the x-axis (coarse Top-$K_2$ stage) and tied $m_1{=}m_3$ along the y-axis (centroid selection and re-ranking). Each cell reports the Recall@100 difference relative to IVF on NQ. The trends are consistent across both encoders and $N_{\mathrm{nprobe}}$. Keeping the precision-critical stages narrow ($m_1{=}m_3 \in \{1,2,4\}$) maintains accuracy close to IVF while allowing the second stage to run with a larger $m_2$. With MiniLM, the gap stays within about 3.5\% at $N_{\text{nprobe}}{=}64$ and within about $2$\% at $N_{\text{nprobe}}{=}128$; with mpnet, it is within about $1$\% for both settings. However, once $m_1, m_3$ approach 16, the degradation becomes substantial. For MiniLM in particular, the Recall@100 drop exceeds 10\%, reflecting that overly wide sensing in the precision-critical stages severely harms fidelity.



Conversely, results show that the second stage is comparatively insensitive: once $m_1$ and $m_3$ are modest, increasing $m_2$ to $8$ has little recall degradation, enabling aggressive mid-stage parallelism. These results support an asymmetric configuration---small $m_1$ and $m_3$ for fidelity, large $m_2$ in the second stage for throughput---leveraging the array’s inherent parallelism without sacrificing recall.




\subsection{Benefit of DTS vs. UBS+LBS}



Fig.~\ref{fig:dts_benefit}(a) and (b) show the advantage of DTS under DTS141 and DTS441, respectively, where DTS applies a complement-based tighter bound than the lenient UBS + LBS scheme as discussed in Section~\ref{sec:ddbam}.
When $m=1$, the two schemes become equivalent (DTS = UBS+LBS), which explains why using $m=1$ in the first and third stages results in negligible differences. However, increasing only $m_1$ to 4 immediately enlarges the recall gap between the two schemes by more than 4\%, and this gap continues to grow as $m$ increases. These results demonstrate that our proposed DTS, which implements a complement-based dual-bound sensing scheme, performs a more precise search than the lenient UBS+LBS approach. 

\vspace{-0.1cm}



\subsection{Noise-Aware Circuit and System Validations}\label{alphasensitivity}

\begin{figure}[t]
    \centering
    \includegraphics[width=0.95\linewidth]{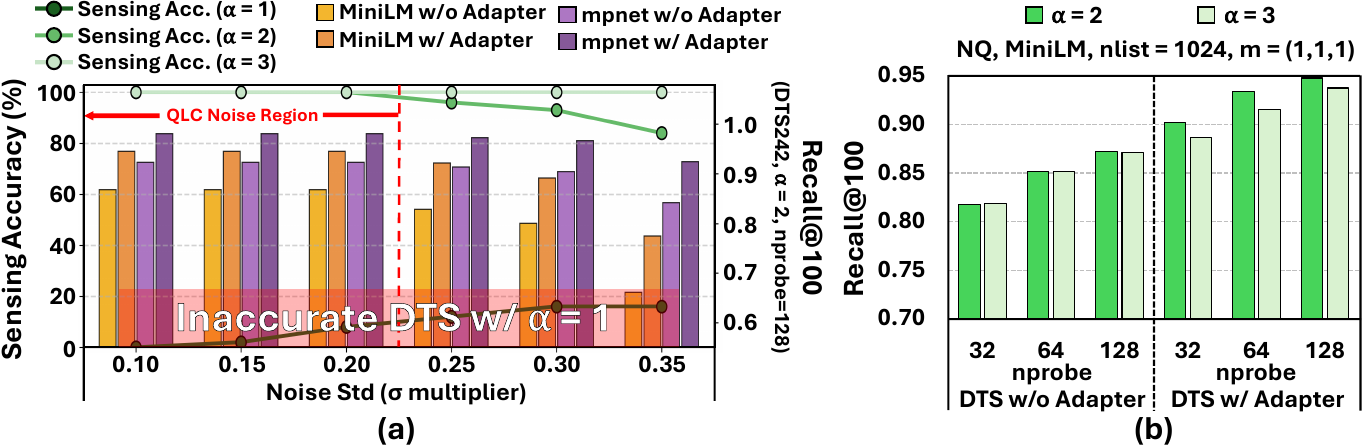}
    \caption{(a) NAND sensing accuracy in QLC and Recall@100 under noise on NQ, and (b) Recall@100 at $\alpha{=}2$ and $3$.}
    \label{fig:sensing}
\end{figure}

\begin{figure}[t]
    \centering
    \includegraphics[width=0.95\linewidth]{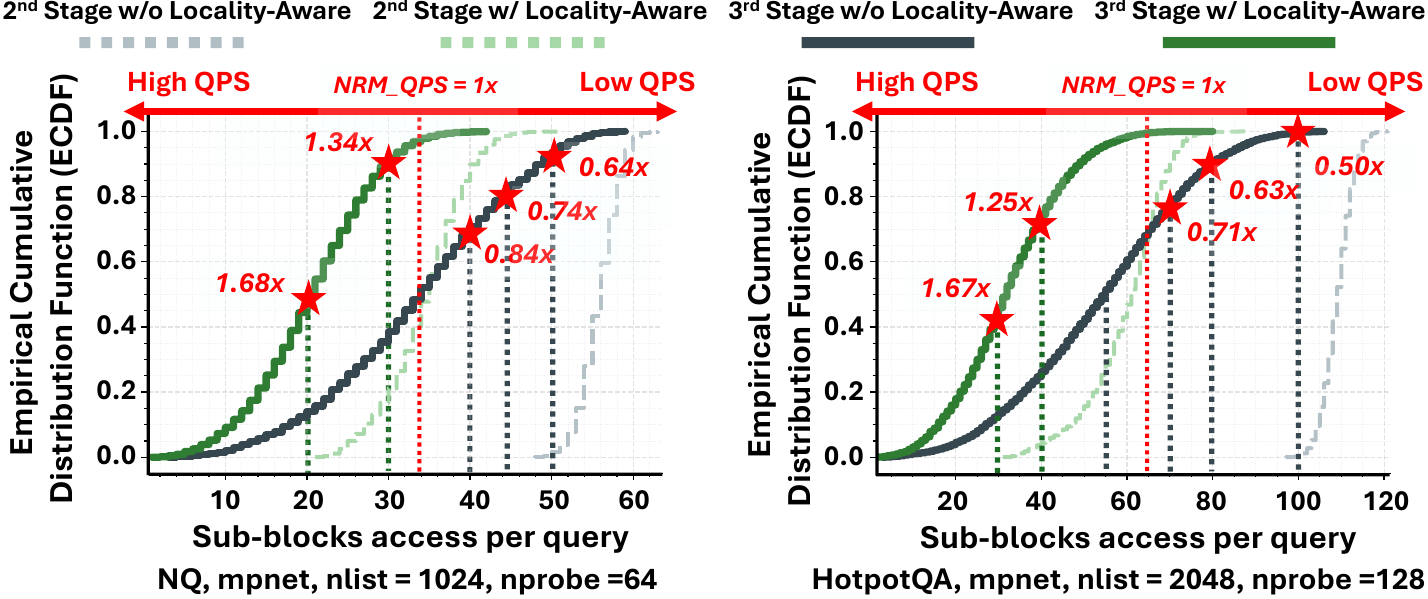}
    \caption{ECDF and normalized QPS of sub-block accesses per query for normal vs. locality-aware mapping. 
    }
    \label{fig:locality}
\end{figure}

\begin{figure*}[t]
    \centering
    \includegraphics[width=0.96\linewidth]{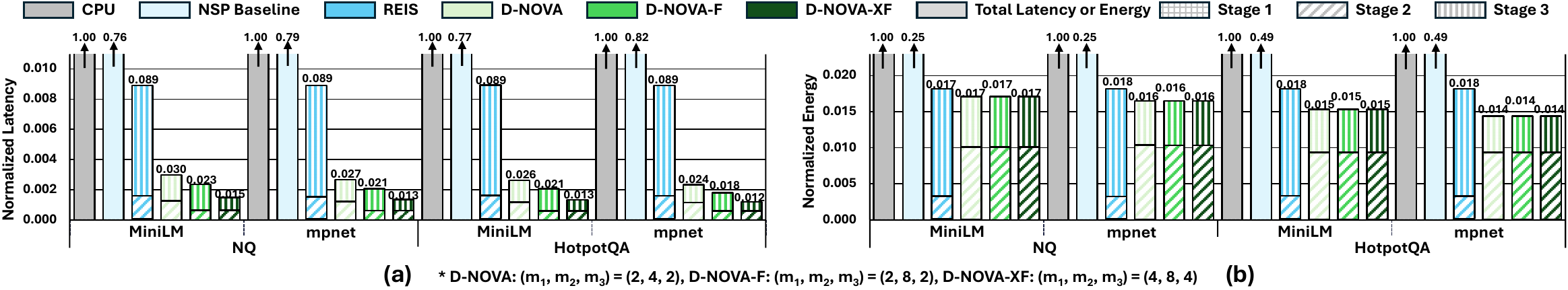}
    \caption{Normalized (a) latency and (b) energy breakdown of D-NOVA compared with CPU, NSP, and REIS baselines.}
    \label{fig:energy_latency}
\end{figure*}

\begin{figure}[t]
    \centering
    \includegraphics[width=0.94\linewidth]{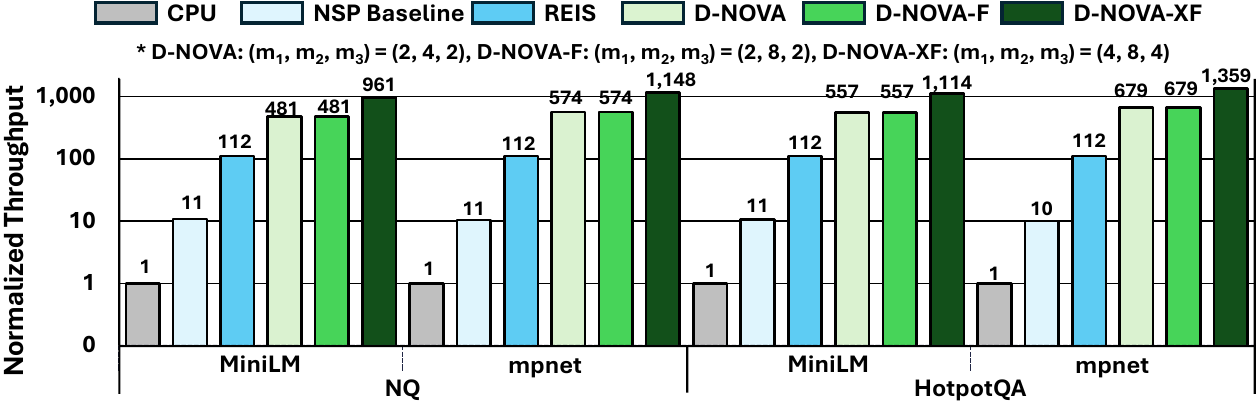}
    \caption{Normalized throughput of D-NOVA vs. baselines.} 
    \label{fig:qps}
\end{figure}

Fig.~\ref{fig:sensing}(a) illustrates the binary sensing accuracy (i.e., UBS-like bound checks) under varying $V_{TH}$ noise. 
We sweep the standard deviation ($\sigma$) of Gaussian $V_{TH}$ noise and perform 1,000 Monte Carlo trials per $\sigma$, covering a wide QLC noise regime ($0.1 \le \sigma \le 0.35$) centered around $\sigma \approx 0.2$, which corresponds to high-RBER conditions ($\sim\!10^{-2}$) observed in state-of-the-art 3D QLC NAND~\cite{papandreou2020open}.
In the critical QLC region, $V_{TH}$ drift causes the sensing accuracy for $\alpha=1$ to collapse below 10\%, rendering DTS ineffective. In contrast, margins of $\alpha \ge 2$ successfully compensate for these variations, restoring sensing reliability across the targeted noise range.  Fig.~\ref{fig:sensing}(b) summarizes the impact of the DTS margin $\alpha$ on Recall@100 for NQ with MiniLM under DTS111. We treat $\alpha{=}2$ as the nominal operating point. Without the adapter, $\alpha{=}3$ stays within approximately $0.1\%$ of the reference accuracy. With the adapter, overall Recall@100 improves, and tighter bounds become more favorable, as the embedding is adapted to the DTS sensing metric. As a result, $\alpha{=}2$ offers a robust trade-off: the bound is tight enough to suppress false matches while remaining tolerant to realistic threshold variation. Although a larger margin ($\alpha{=}3$) remains accurate, it begins to admit additional false positives, particularly at smaller $N_{\text{nprobe}}$. Therefore, we use $\alpha{=}2$ as the default setting for D-NOVA. This sweep shows that DTS is hardened against QLC retention/drift-like threshold broadening through the tolerance margin $\alpha$, which acts as a voltage-domain guardband during bound checking, and provides a robustness/selectivity guideline for margin selection without admitting excessive false positives.


\vspace{-0.2cm}

\subsection{Benefit of Locality-Aware Mapping} \label{sec:la}
Building on Section~\ref{lamapping}, Fig.~\ref{fig:locality} illustrates how the mapping changes the distribution of sub-block activations per query for the second and third stage on NQ and HotpotQA with the mpnet. Each curve is an ECDF over queries, so a leftward shift means that a typical query touches fewer sub-blocks and requires fewer plane activations, directly implying higher QPS. On NQ, locality-aware mapping reduces the median footprint by $\sim$$38\,\%$ in both the second and third stages, so the same IVF index is retrieved from a much tighter subset of the array. On HotpotQA, the median number of sub-block accesses reduces by $\sim$$44\,\%$ in the second and $\sim$$42\,\%$ in the third stage.
Under incremental database updates, this locality benefit is still preserved. Even if an additional 25\% of the database is appended using naive placement, the median sub-block-access reduction remains about 30.4\% on NQ and 33.6--35.2\% on HotpotQA.
Therefore, locality-aware mapping confines the second- and third-stage accesses to fewer sub-blocks and planes, reducing the number of activations per query. Importantly, DTS performs fully parallel sensing within a plane, so concentrating vectors into fewer sub-blocks does not increase per-access latency. While repeated accesses may introduce temporal contention, cluster placement across planes distributes accesses and mitigates such effects.

\vspace{-0.1cm}

\subsection{E2E Latency, Energy, and Throughput}
\label{sec:system_results}
We evaluate three D-NOVA variants with DTS configurations: D-NOVA (DTS242), D-NOVA-F (DTS282), and D-NOVA-XF (DTS484). As shown in Section~\ref{sec:result_m}, DTS484 maintains Recall@100 within a few percent of the IVF baseline, so we include D-NOVA-XF as an aggressive throughput-oriented design while preserving accuracy.
On the other hand, D-NOVA is the conservative variant, achieving high accuracy despite low parallelism.

\noindent\textbf{Latency \& Energy Improvement:} Fig.~\ref{fig:energy_latency}(a) and (b) summarize end-to-end latency and per-query energy, normalized to the CPU baseline. Across datasets and encoders, all D-NOVA variants deliver the lowest latency and energy. Even the conservative D-NOVA configuration reduces latency by $30\times$ and energy by $15$--$34\times$ relative to NSP, achieves $33$--$41\times$ lower latency and $58$--$71\times$ lower energy than the CPU software baseline, and further reduces latency by $3.01$--$3.79\times$ (D-NOVA), $3.80$--$4.99\times$ (D-NOVA-F), and $5.98$--$7.53\times$ (D-NOVA-XF) with additional energy savings of $6$--$27\%$ over REIS.
These results directly quantify the benefit of avoiding off-array final re-ranking. D-NOVA keeps all embeddings in flash, performs both coarse and fine scoring through in-array DTS-based binary sensing, and transfers only compact scores/metadata to the controller, avoiding full multi-level QLC candidate readout and controller-side INT8 re-ranking. In contrast, REIS is the closest baseline for this case: it performs the earlier NAND-side binary search stages, but still moves tens of thousands of QLC-resident candidate embeddings off-array to the SSD controller for final INT8 re-ranking. NSP represents a more general near-storage design that reads INT8 embeddings from flash before external similarity computation. Eliminating these off-array candidate accesses and processor computations removes the dominant sources of latency and energy in prior designs.
Because all D-NOVA variants activate nearly the same set of WLs for a given query, varying $m$ primarily reshapes the sensing schedule (e.g., sequential vs. parallel) without changing the total number of WL activations. As WL toggling dominates overall energy consumption, the total per-query energy remains largely similar across different $m$ configurations.
Beyond eliminating re-ranking-time candidate readout, D-NOVA further reduces energy by minimizing sub-block activations through locality-aware mapping. Overall, although per-sensing energy is similar, D-NOVA achieves much faster end-to-end latency and far lower total energy by requiring fewer sensing cycles and eliminating almost all off-array DRAM and compute overhead.

\noindent \textbf{Throughput Improvement:} Fig.~\ref{fig:qps} shows the corresponding steady-state throughput (QPS) under large-batch operation, normalized to the CPU baseline. With the pipeline saturated, D-NOVA and D-NOVA-F sustain roughly $481$--$679\times$ of the CPU QPS and about $42.7$--$67.9\times$ of the NSP baseline. Relative to REIS, D-NOVA and D-NOVA-F improve throughput by about $4.3$--$6.1\times$, while D-NOVA-XF reaches $8.6$--$12.1\times$ higher QPS. In this configuration, D-NOVA and D-NOVA-F exhibit similar QPS because the larger $m_2$ in D-NOVA-F shortens the coarse second stage, so the third stage (with the same $m_3$ in both) becomes the dominant bottleneck. This trade-off can be re-tuned per deployment by shrinking $K_1$ or adjusting $N_{\text{nprobe}}$ and $m_2$ to shift more work into the highly parallel second stage while meeting the target accuracy.

\vspace{-0.35cm}

\subsection{Overhead of D-NOVA System} \label{sec:overhead}

DTS is evaluated by page-buffer comparators that reuse existing SA front-ends in binary pass/fail mode.
To support delta-based accumulation, D-NOVA introduces a lightweight shared accumulator in the peripheral logic. 
Our synthesized design shows that sharing one accumulator across 256 BLs (64 accumulators/plane) results in a total area of 0.023~mm$^2$ per plane, including multiplexer, corresponding to <1.5\% of the 1.476~mm$^2$ plane area~\cite{hsu2023storage}.
This overhead is amortized within existing CMOS-under-Array (CuA) regions~\cite{parat2015floating, noh2025flexible, 10.1145/3538643.3539742}, resulting in negligible die-level impact.
The accumulation is tightly pipelined with DTS: while SL-level sensing is performed, CL outputs from previously sensed WLs are forwarded to the accumulator, allowing score updates to proceed in parallel with ongoing sensing.
The accumulation latency (256 BLs @ 1~GHz $\approx 0.25~\mu$s) is negligible compared to sensing ($\approx 30~\mu$s) and is fully hidden by pipelined execution.
The BL grouping (e.g., 256 per accumulator) is chosen to balance accumulator fan-in and multiplexer complexity, avoiding long critical paths and excessive routing overhead in the peripheral logic.
Detailed configurations are provided in Section~\ref{sec:dnova_config}.

In addition, D-NOVA leverages multi-WL activation to increase intra-string parallelism. 
To enable $m$-adaptive sensing, D-NOVA leverages existing per-WL voltage control, where each WL already supports $V_{\mathrm{read}}$ and $V_{\mathrm{pass}}$ during standard reads~\cite{11367431,9366003} along with existing multi-WL activation modes~\cite{park2022flash, Lee2015,Sharon2014, jp2003222422a, salehi2022memory}. We reuse the same WL drivers for multi-WL activation.
Enabling multiple WLs can increase the effective on-resistance of the string, potentially leading to increased latency. Using our in-house DTS circuit simulator, we compare the BL discharge time against the reference $t_{\mathrm{DISCH}}$ reported in~\cite{10.1145/3445814.3446719} and observe that even with $m=8$, the sensing latency differs by only 5.6\%. Since $t_{\mathrm{DISCH}}$ constitutes only a small portion of the overall read latency budget, this difference translates to negligible impact on end-to-end read performance, indicating that the practical overhead of multi-WL activation in our operating regime is minimal. Prior work~\cite{park2022flash} further confirms that activating up to 48 WLs is feasible within a read timing margin of 24~$\mu$s, based on validation across multiple dies of real silicon.

\section{Conclusion}

D-NOVA demonstrates the feasibility of executing an entire retrieval pipeline fully inside the storage device, eliminating the memory bandwidth bottleneck at the system level. We introduce a novel Dual-Bound Tight Similarity Sensing (DTS) mechanism integrated into 3D NAND flash and a lightweight contrastive adapter, trained once \textit{offline}, that aligns embeddings to this discrete metric, thereby preserving high recall accuracy. This hardware–software co-designed in-storage architecture achieves $4.3$–$12.1\times$ higher throughput and $1.1$–$1.26\times$ lower energy than the state-of-the-art in-storage accelerator~\cite{reis}, with feasibility validated through detailed circuit-level noise simulations and overhead analysis. These results establish a new direction for efficient and scalable RAG systems.




\begin{acks}
This work was supported in part by the Center for Processing with Intelligent Storage and Memory (PRISM) under Semiconductor Research Corporation (SRC) grant 2023-JU-3135 and by CoCoSys, both centers in JUMP 2.0, an SRC program sponsored by DARPA. This work was also supported in part by NSF grants \#2112665, \#2211386, \#2052809, and \#2112167.
\end{acks}

\bibliographystyle{ACM-Reference-Format}
\bibliography{References}

\end{document}